\documentclass[aps,prl,reprint,groupedaddress]{revtex4-1}
\usepackage{graphicx} 
\usepackage{dcolumn} 
\usepackage{bm}
\usepackage{hyperref}

\begin{document}

\title{Nuclear dynamics and particle production near threshold energies in heavy-ion collisions}
\thanks{Supported by the National Natural Science Foundation of China (Nos. 11675226 and 11722546) and the Major State Basic Research Development Program of China (Nos. 2014CB845405 and 2015CB856903)}

\author{FENG Zhao-Qing}
\affiliation{Institute of Modern Physics, Chinese Academy of Sciences, Lanzhou 730000, China}
\email{Corresponding author: fengzhq@impcas.ac.cn}

\begin{abstract}
Recent progress of the quantum molecular dynamics model for describing the dynamics of heavy-ion collisions is viewed, in particular the nuclear fragmentation, isospin physics, particle production and in-medium effect, hadron-induced nuclear reactions, hypernucleus etc. The neck fragmentation in Fermi-energy heavy-ion collisions is investigated for extracting the symmetry energy at subsaturation densities. The isospin effects, in-medium properties and the behavior of high-density symmetry energy in medium and high energy heavy-ion collisions are thoroughly discussed. The hypernuclide dynamics formed in heavy-ion collisions and in hadron induced reactions is analyzed and addressed in the future experiments at the High-Intensity heavy-ion Accelerator Facility (HIAF).
\end{abstract}

\keywords{LQMD transport model, Heavy-ion reactions, Hadron induced nuclear reactions, Symmetry energy, Strangeness production}

\maketitle

\section{Introduction}

The nuclear dynamics in heavy-ion collisions and in hadron induced reactions at the medium and high energies is complicated and associated with the many-body interaction, collision dynamics, fermionic nature etc. Over past several decades, two sorts of models based on transport theories have been developed for investigating the hot and dense nuclear matter created in heavy-ion collisions, i.e., the quantum molecular dynamics and Boltzmann Uehling Uhlenbeck (BUU) transport models \cite{Ue33,Be88,Ai86,Ai91}. The liquid-gas phase transition formed in dilute nuclear matter and the quark-gluon plasma (QGP) phase transition in dense matter attracted much attention and have obtained progress both in theories and in experiments \cite{Po95,Ma99,Sh93,Ca99,Ad05,So16,Ad17}. The isospin degree of freedom in the nuclear matter plays a significant role on the phase transition.

The nuclear equation of state (EOS) is a basic quantity for describing the nuclear matter with the density and temperature. The EOS in the isospin asymmetric nuclear matter is usually expressed with the energy per nucleon as $E(\rho, T, \delta)=E(\rho, T, 0)+E_{\textrm{sym}}(\rho,T)\delta^{2}+ \mathcal{O}(\delta^{2})$ in terms of baryon density $\rho$, isospin asymmetry $\delta=(\rho_{n}-\rho_{p})/(\rho_{n}+\rho_{p})$ and temperature $T$. For treating the properties of a nucleus and cold dense matter, the temperature is usually not taken into account. The high-density symmetry energy $E_{\textrm{sym}}$ is a key ingredient in understanding the stellar structure, the cooling of protoneutron stars, the nucleosynthesis during supernova explosion of massive stars and even the binary neutron star merger \cite{Ab17,Li17}. The symmetry energy at suprasaturation densities can be constrained from the observables emitted from the high-density regime in nucleus-nucleus collisions. Up to now, the high-density behavior of symmetry energy is not well understood. With operating and constructing the new generation radioactive beam facilities in the world, such as the Cooling Storage Ring (Institute of Modern Physics, Lanzhou, China), Facility for Antiproton and Ion Research (GSI, Darmstadt, Germany), RIKEN (Japan), SPIRAL2 (GANIL, Caen, France), Facility for Rare Isotope Beams (MSU, East Lansing, USA), the isospin physics is to be attracted much attention both in theories and in experiments. Besides nucleonic observables, particles produced in heavy-ion collisions would be preferable probes for extracting the information of high-density phase diagram. Kaons as probing the high-density EOS were proposed for the first time \cite{Ai85}. The available experimental data from KaoS collaboration for $K^{+}$ production favored a soft EOS at high baryon densities associated with transport model calculations \cite{St01,Li95b,Fu01,Ha06a}. Similar structure for $\Lambda$ production on the EOS was found in Ref.\cite{Fe11a}. The ratios of isospin particles produced in heavy-ion collisions such as $\pi^{-}/\pi^{+}$, $K^{0}/K^{+}$, $\Sigma^{-}/\Sigma^{+}$ etc \cite{Li02,Li05,Fe06,To10,Pr10,Fe10a,Fe10b}, the yields of $\eta$ etc \cite{Yo13,Ch17}, and the flow difference of isospin particles \cite{Gi10,Fe12a} have been proposed as sensitive probes for extracting the high-density behavior of the nuclear symmetry energy.

Properties of hadrons in dense nuclear matter is related to the Quantum Chromodynamics (QCD) phase structure, in particular the chiral symmetry restoration, phase transition from hadronic to partonic degrees of freedom, hypernucleus formation, nuclear EOS etc \cite{Gi95,Fr07,To11}. Nucleus-nucleus collisions provide possibilities for exploring the in-medium properties of hadrons, which are related to the interaction potentials and production cross sections in nuclear reactions. The energy spectra of particles are distorted by the surrounding nucleons. It has obtained progress in extracting the in-medium properties of hadrons in dense nuclear matter, in particular for strange particles $K$, $\overline{K}$, $\Lambda$ and $\Sigma$ \cite{Li97,Li98,Fu06,Ha12}. Studies of hypernuclei attract much attention over the past several decades. The interested topics related to hypernuclei are the hyperon-nucleon and hyperon-hyperon interactions, opening a new horizon with strangeness (three-dimensional nuclear chart) in nuclear physics and probing the in-medium properties of hadrons and the inner structure of a nucleus \cite{Gi95, Ha06}. Moreover, hyperons as essential ingredients impact the EOS at high-baryon densities. The strangeness constitution in dense matter softens the high-density EOS, and consequently decreases the mass of neutron stars\cite{Ji13,Ji17,We12}. Since the first observation of $\Lambda$-hypernuclide in nuclear multifragmentation reactions induced by cosmic rays in 1950s \cite{Da53}, a remarkable progress has been obtained in producing hypernuclides via different reaction mechanism, such as hadron (pion, kaon, proton) induced reactions, bombarding a fixed target with high-energy photons or electrons, and fragmentation reactions with high energy heavy-ion collisions. Experimental collaborations of nuclear physics facilities in the world, e.g., PANDA \cite{Pand}, FOPI/CBM and Super-FRS/NUSTAR \cite{Sa12} at FAIR (GSI, Germany), STAR at RHIC (BNL,USA) \cite{Star}, ALICE at LHC (CERN) \cite{Do13}, NICA (Dubna, Russia) \cite{Nica}, J-PARC (Japan) \cite{Ta12}, HIAF (IMP, China) \cite{Ya13} have started or planned to investigate hypernuclei and their properties. In the laboratories, strangeness nuclear physics is to be concentrated on the isospin degree of freedom (neutron-rich/proton-rich hypernuclei), multiple or anti-strangeness (s=-2 or s=1), high-density hypermatter production etc. A more localized energy deposition enables the secondary collisions available for producing hyperons in antiproton-nucleus collisions. The cold quark-gluon plasma (QGP) could be formed from the annihilation of antiprotons on nuclei \cite{Ra80,Ra88} and the searching is still undergoing in experiments.

In this article, I will review the recent progress on the theoretical treatment of nuclear dynamics in heavy-ion collisions. The symmetry energy, strange particles and in-medium effects, and hyperfragment formation are to be discussed. In section II, I will give a transport model description for simulating the nuclear dynamics essentially based on the quantum molecular dynamics (QMD). The nuclear dynamics and isospin effect in heavy-ion collisions are shown in section III. The hadron induced reactions and hypernuclide production are investigated in section IV. Summary and perspective on the medium-energy nuclear dynamics are presented in section V.

\section{The quantum molecular dynamics model and its modification}

In the QMD transport model, the $i$th nucleon is represented with a Gaussian wave packet \cite{Ai86,Ai91}
\begin{equation}
\psi_{i}(\mathbf{r},t)=\frac{1}{(2\pi  \sigma_{r}^{2})^{3/4}} \exp\left[-\frac{(\mathbf{r}-\mathbf{r}_{i}(t))^{2}}{4\sigma_{r}^{2}}\right]\cdot
\exp\left(\frac{i\mathbf{p}_{i}(t)\cdot\mathbf{r}}{\hbar}\right).
\end{equation}
Here $\mathbf{r}_{i}(t)$, $\mathbf{p}_{i}(t)$ are the centers of the wave packet in the coordinate and momentum space, respectively. The $\sigma_{r}$ is the width of the Gaussian wave packet. The $\emph{N}$-body wave function is assumed as the direct product of the coherent states $\Phi(\mathbf{r},t)=\prod_{i}\psi_{i}(\mathbf{r},t)$ and the anti-symmetrization is neglected. Therefore, the fermionic nature in the dynamical evolution is lost. After performing Wigner transformation, the phase-space density is obtained as
\begin{equation}
f(\mathbf{r},\mathbf{p},t)=\sum_{i}f_{i}(\mathbf{r},\mathbf{p},t),
\end{equation}
\begin{eqnarray}
f_{i}(\mathbf{r},\mathbf{p},t)= && \frac{8}{h^{3}}\exp\left[-\frac{(\mathbf{r}-\mathbf{r}_{i}(t))^{2}}{2\sigma_{r}^{2}}\right]
\nonumber  \\
&&   \times \exp\left[-\frac{2(\mathbf{p}-\mathbf{p}_{i}(t))^{2}\cdot \sigma_{r}^{2}}{\hbar^{2}}\right].
\end{eqnarray}
The density distributions $\rho(\mathbf{r},t)=\sum_{i}\rho_{i}(\mathbf{r},t)$ in the coordinate and $g(\mathbf{p},t)=\sum_{i}g_{i}(\mathbf{p},t)$ in the momentum space can be easily evaluated with the phase-space density as
\begin{eqnarray}
\rho_{i}(\mathbf{r},t)=\frac{1}{(2\pi)^{3/2}\sigma_{r}^{3}}\exp\left[-\frac{(\mathbf{r}-\mathbf{r}_{i}(t))^{2}}{2\sigma_{r}^{2}}\right], \\
g_{i}(\mathbf{p},t)    =(8\pi)^{3/2}(\sigma_{r}/h)^{3}\exp\left[-\frac{2(\mathbf{p}-\mathbf{p}_{i}(t))^{2}\cdot \sigma_{r}^{2}}{\hbar^{2}}\right].
\end{eqnarray}

The QMD transport model was improved by different groups, such as implementing the Pauli potential into the mean-field potential \cite{Pe92}, extending to the relativistic heavy-ion collisions (TuQMD, UrQMD) \cite{Fu06,Bl99}, introducing the cooling method in the initialization and the time-dependent wave-packet width (EQMD) \cite{Ma90,Ma98}, distinguishing the isospin degree of freedom (IQMD) \cite{Ha98,Ch98}, taking the phase-space constraint approach into the dynamical evolution and nucleon-nucleon collisions (CoMD) \cite{Pa01}. Recently, the ImQMD model by the CIAE group was proposed for describing the fusion-fission, quasifission dynamics, multinucleon transfer, symmetry energy etc \cite{Wa02,Wa14,Ti08,Zh13,Zh14}. The mean-field potential in the ImQMD model is derived selfconsistently from the Skyrme energy-density functional. The shell effect in the fusion reactions was investigated within the framework of the QMD model by the Lanzhou group and a phenomenological formula was proposed \cite{Fe05,Fe08}. Recently, the giant dipole resonance formed in heavy-ion collisions was investigated by SINAP group for probing the $\alpha$ clustering structure in a nucleus and the EOS within the QMD model \cite{He14,Ta13}.

In this work, I will give a description of the quantum molecular dynamics model developed by the Lanzhou group (LQMD). The dynamics of the resonances ($\Delta$(1232), N*(1440), N*(1535), etc), hyperons ($\Lambda$, $\Sigma$, $\Xi$, $\Omega$) and mesons ($\pi$, $\eta$, $K$, $\overline{K}$, $\rho$, $\omega$) is described via hadron-hadron collisions, Pauli-blocking, decays of resonances, mean-field potentials, and corrections on threshold energies of elementary cross sections \cite{Fe13a,Fe15a}. Besides the hadron-hadron collisions, we have further included the annihilation channels, charge-exchange reaction, elastic and inelastic collisions in antinucleon-nucleon collisions for understanding antiproton induced reactions \cite{Fe14a}. The LQMD model has been applied to investigate the medium and high energy heavy-ion collisions, antiproton induced reactions, meson-nucleus dynamics etc.

In the LQMD model, the time evolutions of the baryons (nucleons and resonances) and mesons in the reaction system are governed by Hamilton's equations of motion under the self-consistently generated mean-field potential as
\begin{eqnarray}
\dot{\mathbf{p}}_{i}=-\frac{\partial H}{\partial\mathbf{r}_{i}},
\quad \dot{\mathbf{r}}_{i}=\frac{\partial H}{\partial\mathbf{p}_{i}}.
\end{eqnarray}
The Hamiltonian of baryons consists of the relativistic energy, the effective potential and the momentum dependent interaction. The effective potential is composed of the Coulomb interaction and the local potential
\begin{equation}
U_{int}=U_{Coul}+U_{loc}.
\end{equation}
The Coulomb interaction potential is written as
\begin{equation}
U_{Coul}=\frac{1}{2}\sum_{i,j,j\neq i}\frac{e_{i}e_{j}}{r_{ij}}erf(r_{ij}/2\sigma_{r})
\end{equation}
where the $e_{j}$ is the charged number including protons and charged resonances. The $r_{ij}=|\mathbf{r}_{i}-\mathbf{r}_{j}|$ is the relative distance of two charged particles.

The local potential can be derived from the energy-density functional as
\begin{equation}
U_{loc}=\int V_{loc}(\rho(\mathbf{r}))d\mathbf{r},
\end{equation}
which is constructed with the 2-body and 3-body Skyrme-type interaction as
\begin{equation}
U_{loc}^{2}=\frac{a_{2}}{2}\sum_{i,j,j\neq i}\rho_{i}(\mathbf{r}) \delta(\mathbf{r}-\mathbf{r}^{\prime})\rho_{j}(\mathbf{r}^{\prime})d\mathbf{r}d\mathbf{r}^{\prime},
\end{equation}
and
\begin{eqnarray}
U_{loc}^{3}= && \frac{a_{3}}{6}\sum_{i,j,k,j\neq i,j\neq k, i\neq k}\rho_{i}(\mathbf{r})\rho_{j}(\mathbf{r}^{\prime})\rho_{k}(\mathbf{r}^{\prime\prime}) \delta(\mathbf{r}-\mathbf{r}^{\prime})
\nonumber \\
&&   \times \delta(\mathbf{r}-\mathbf{r}^{\prime\prime}) d\mathbf{r}d\mathbf{r}^{\prime}d\mathbf{r}^{\prime\prime}.
\end{eqnarray}
In the model, the 2-body nucleon-nucleon (NN) interaction includes the attractive potential, symmetry energy, surface term and momentum dependent interaction. The 3-body interaction is embodied in the short-range repulsive force and an approximation treatment is taken in evaluating Eq. (11) \cite{Ai91}. The energy-density functional is expressed by
\begin{eqnarray}
V_{loc}(\rho)=&& \frac{\alpha}{2}\frac{\rho^{2}}{\rho_{0}} +
\frac{\beta}{1+\gamma}\frac{\rho^{1+\gamma}}{\rho_{0}^{\gamma}} + E_{sym}^{loc}(\rho)\rho\delta^{2}
\nonumber \\
&& + \frac{g_{sur}}{2\rho_{0}}(\nabla\rho)^{2} + \frac{g_{sur}^{iso}}{2\rho_{0}}[\nabla(\rho_{n}-\rho_{p})]^{2},
\end{eqnarray}
where the $\rho_{n}$, $\rho_{p}$ and $\rho=\rho_{n}+\rho_{p}$ are the neutron, proton and total densities, respectively, and the $\delta=(\rho_{n}-\rho_{p})/(\rho_{n}+\rho_{p})$ being the isospin asymmetry. The surface coefficients $g_{sur}$ and $g_{sur}^{iso}$ are taken as 23 MeV fm$^{2}$ and -2.7 MeV fm$^{2}$, respectively.
The bulk parameters $\alpha$, $\beta$ and $\gamma$ are readjusted after inclusion the momentum term in order to reproduce the compression modulus of symmetric nuclear matter and the binding energy of isospin symmetric nuclear matter at saturation density. The $E_{sym}^{loc}$ is the local part of the symmetry energy, which can be adjusted to mimic predictions of the symmetry energy calculated by microscopical or phenomenological many-body theories and has two-type forms as follows:
\begin{equation}
E_{sym}^{loc}(\rho)=\frac{1}{2}C_{sym}(\rho/\rho_{0})^{\gamma_{s}},
\end{equation}
and
\begin{equation}
E_{sym}^{loc}(\rho)=a_{sym}(\rho/\rho_{0})+b_{sym}(\rho/\rho_{0})^{2}.
\end{equation}
The parameters $C_{sym}$, $a_{sym}$ and $b_{sym}$ are taken as the values of 52.5 MeV, 43 MeV, -16.75 MeV and 23.52 MeV, 32.41 MeV, -20.65 MeV corresponding to the mass splittings of $m_{n}^{\ast}>m_{p}^{\ast}$ and $m_{n}^{\ast}<m_{p}^{\ast}$, respectively. The parameter $C_{sym}$ is taken as the value of 38 MeV for the case without the mass splitting. The values of $\gamma_{s}$=0.5, 1., 2. have the soft, linear and hard symmetry energy, respectively, and the last formula corresponding to a supersoft symmetry energy.

\begin{table*}
\caption{The parameters and properties of isospin symmetric EoS used in the LQMD model at the density of 0.16 fm$^{-3}$.} \vspace*{-10pt}
\begin{center}
\def\temptablewidth{0.9\textwidth}
{\rule{\temptablewidth}{1pt}}
\begin{tabular*}{\temptablewidth}{@{\extracolsep{\fill}}ccccccccc}
&Parameters             &$\alpha$ (MeV)   &$\beta$  (MeV)   &$\gamma$    &$C_{mom}$ (MeV)   &$\epsilon$ (c$^{2}$/MeV$^{2}$)  &$m_{\infty}^{\ast}/m$    &$K_{\infty}$ (MeV)    \\
\hline
&PAR1    &-215.7   &142.4   &1.322   &1.76  &5$\times$10$^{-4}$    &0.75    &230      \\
&PAR2    &-226.5   &173.7   &1.309   &0.       &0.     &1.    &230            \\
\end{tabular*}
{\rule{\temptablewidth}{1pt}}
\end{center}
\end{table*}

 A Skyrme-type momentum-dependent potential is used in the LQMD model \cite{Fe11b}
\begin{eqnarray}
U_{mom}=&& \frac{1}{2\rho_{0}}\sum_{i,j,j\neq i}\sum_{\tau,\tau'}C_{\tau,\tau'}\delta_{\tau,\tau_{i}}\delta_{\tau',\tau_{j}}\int\int\int d \textbf{p}d\textbf{p}'d\textbf{r}   \nonumber \\
&& \times f_{i}(\textbf{r},\textbf{p},t) [\ln(\epsilon(\textbf{p}-\textbf{p}')^{2}+1)]^{2} f_{j}(\textbf{r},\textbf{p}',t).
\end{eqnarray}
Here $C_{\tau,\tau}=C_{mom}(1+x)$, $C_{\tau,\tau'}=C_{mom}(1-x)$ ($\tau\neq\tau'$) and the isospin symbols $\tau$($\tau'$) represent proton or neutron. The parameters $C_{mom}$ and $\epsilon$ was determined by fitting the real part of optical potential as a function of incident energy from the proton-nucleus elastic scattering data. The parameter $x$ is the strength of the isospin splitting of nucleon effective mass, e.g., the value of -0.65 leading to the splitting of $m^{\ast}_{n}>m^{\ast}_{p}$ in nuclear medium. Both sets of the parameters with momentum dependent and independent interactions give the same compression modulus of K=230 MeV for isospin symmetric nuclear matter as shown in Table 1.

\begin{figure}
\includegraphics[width=8 cm]{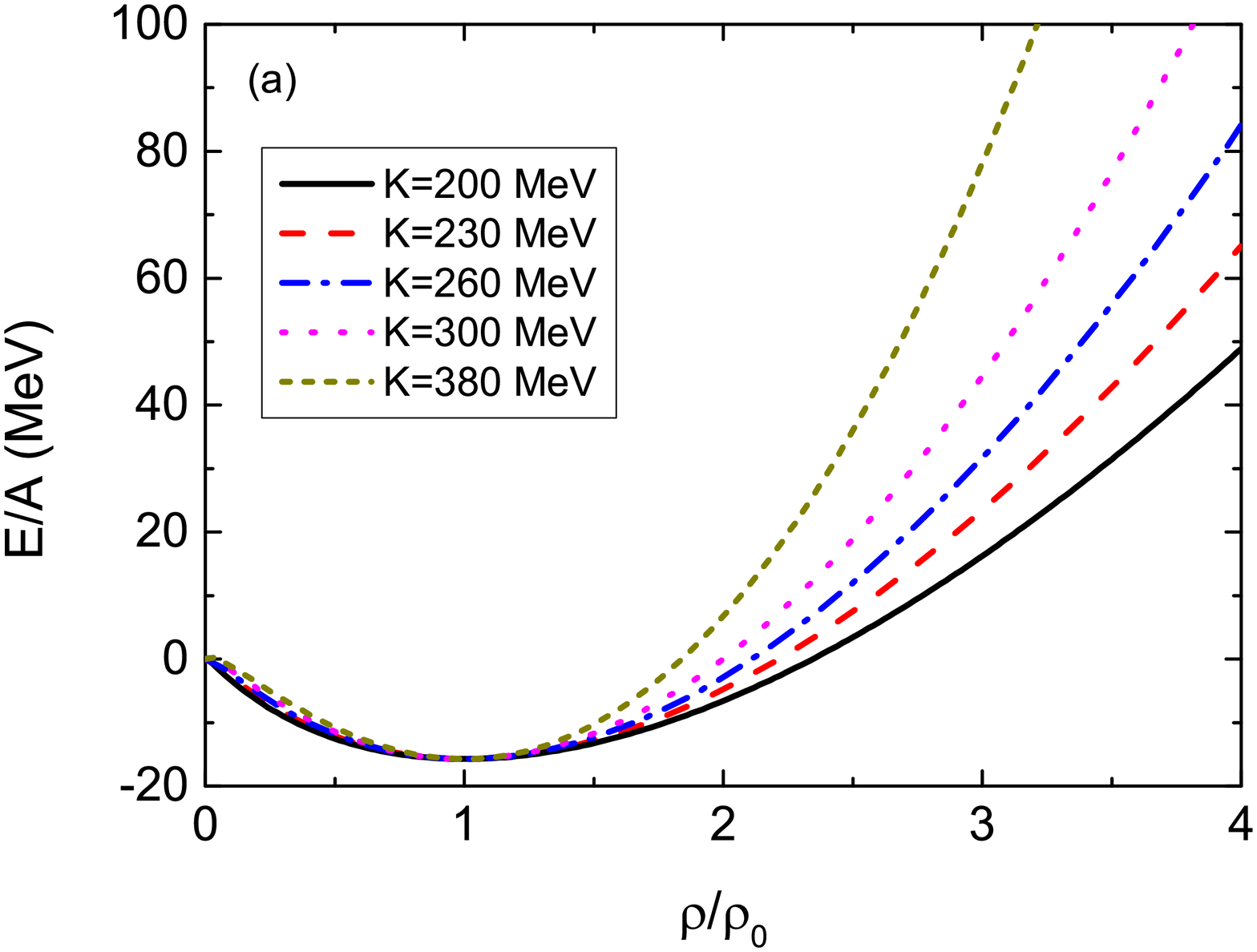}
\includegraphics[width=8 cm]{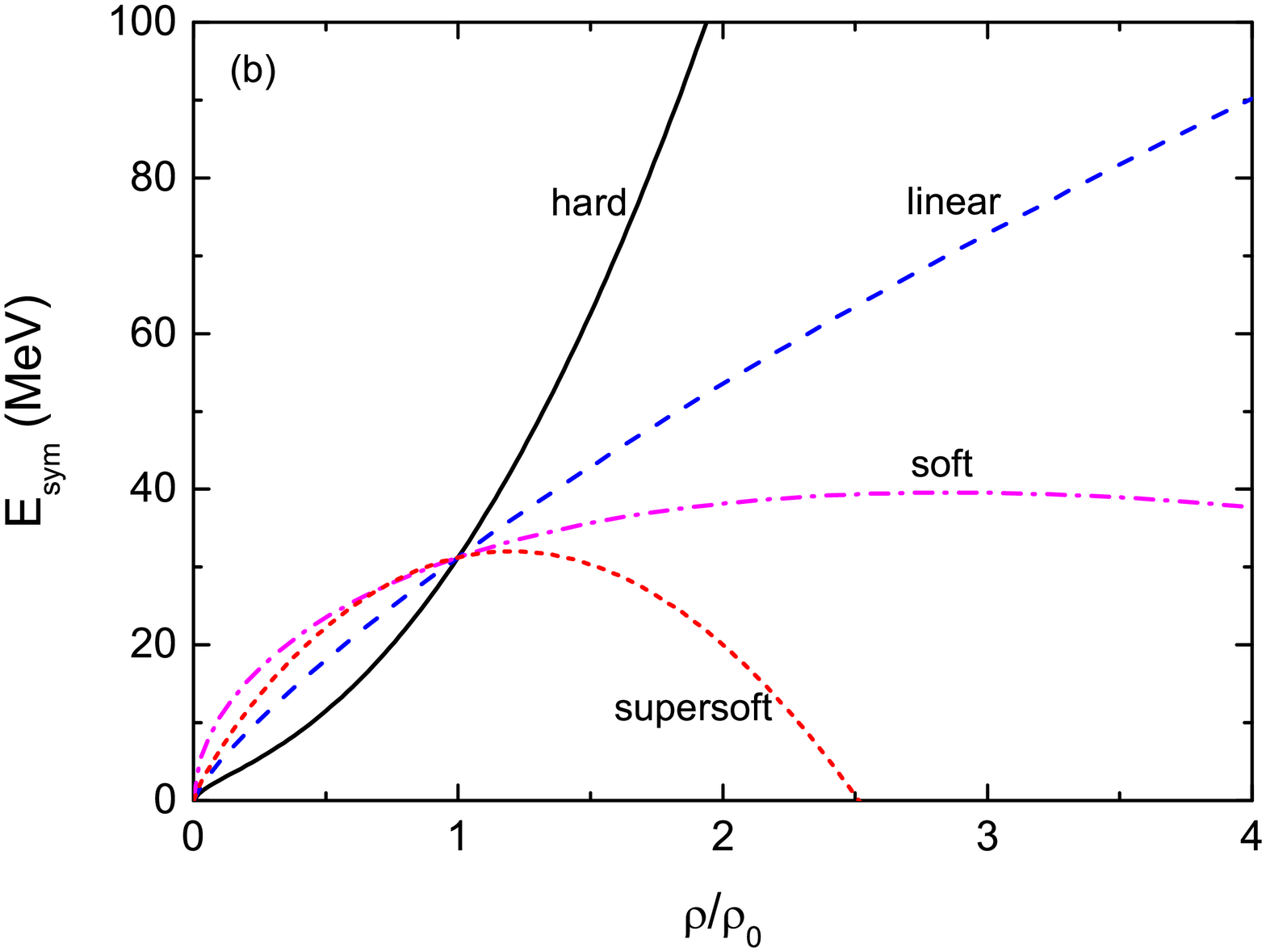}
\caption{(Color online) Density dependence of the symmetric nuclear equation of state at different compression modulus and symmetry energies of supersoft, soft, linear and hard cases.}
\end{figure}

The symmetry energy per nucleon in the LQMD model is composed of three parts, namely the kinetic energy, the local part and the momentum dependence of the potential energy as
\begin{equation}
E_{sym}(\rho)=\frac{1}{3}\frac{\hbar^{2}}{2m}\left(\frac{3}{2}\pi^{2}\rho\right)^{2/3}+E_{sym}^{loc}(\rho)+E_{sym}^{mom}(\rho).
\end{equation}
After an expansion to second order around the normal density, the symmetry energy can be expressed as
\begin{equation}
E_{sym}(\rho) \approx E_{sym}(\rho_{0})+\frac{L}{3}\left(\frac{\rho-\rho_{0}}{\rho_{0}}\right)+\frac{K_{sym}}{18}\left(\frac{\rho-\rho_{0}}{\rho_{0}}\right)^{2}
\end{equation}
in terms of a slope parameter of $L=3\rho_{0}(\partial E_{sym}/\partial \rho)|_{\rho=\rho_{0}}$ and a curvature parameter of $K_{sym}=9\rho_{0}^{2}(\partial^{2} E_{sym}/\partial \rho^{2})|_{\rho=\rho_{0}}$. The values of slope parameters are 203.7 MeV, 124.9 MeV, 85.6 MeV and 74.7 MeV for the hard, linear, soft and supersoft symmetry energies, respectively, and the corresponding 448 MeV, -24.5 MeV, -83.5 MeV and -326 MeV for the curvature parameters. Figure 1 is a comparison of the isospin-symmetric nuclear EOS and nuclear symmetry energy. The EOS of different compression modulus and the symmetry energy of different stiffness cross at the saturation density with the values of -16 MeV and 31.5 MeV, respectively. The difference is pronounced at the high density. The observables from the reaction zone in heavy-ion collisions can extract the high-density EOS.

\begin{figure*}
\includegraphics[width=16 cm]{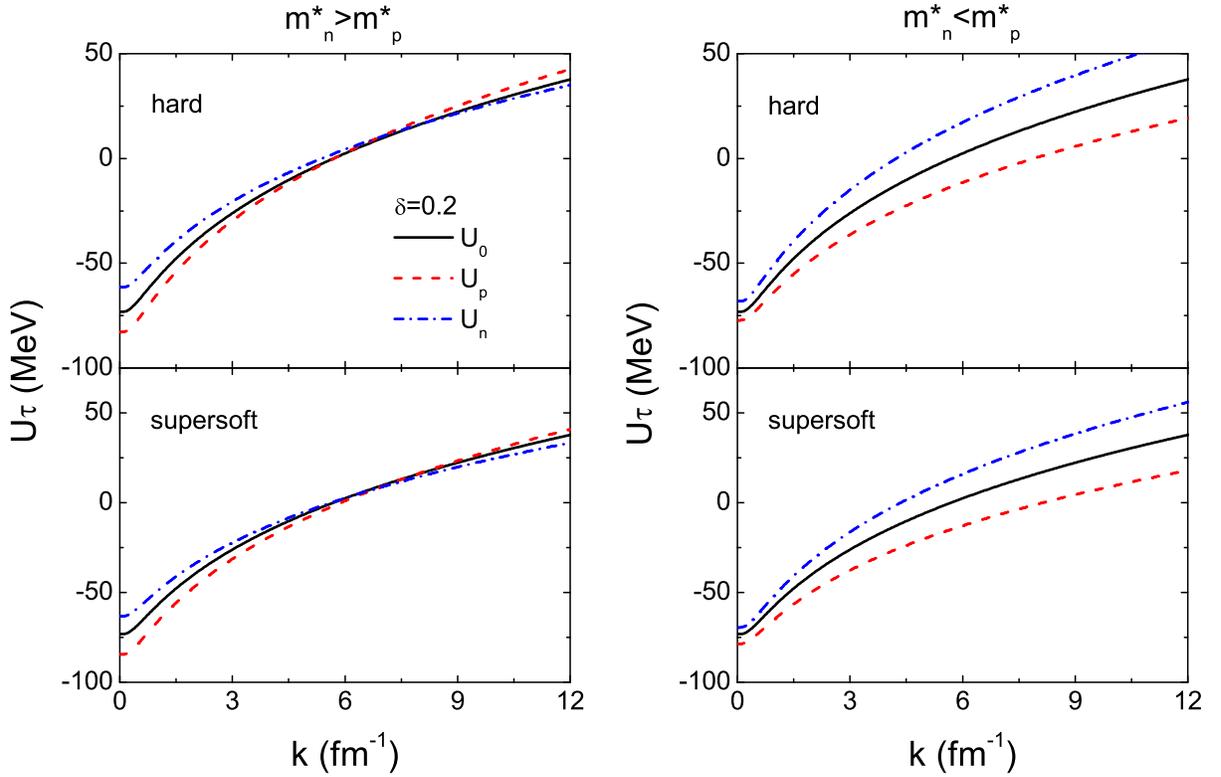}
\caption{(Color online) Momentum dependence of single-nucleon optical potential at the saturation density for isospin symmetric matter with the solid lines ($\delta$=0) and neutron-rich matter ($\delta$=0.2) with the mass splittings of $m_{n}^{\ast}>m_{p}^{\ast}$ (left panel) and $m_{n}^{\ast}<m_{p}^{\ast}$ (right panel), respectively.}
\end{figure*}

The single-particle optical potential can be obtained from the density functional as \cite{Fe12b}
\begin{eqnarray}
U_{\tau}(\rho,\delta,\textbf{p}) = && \frac{1}{\rho_{0}}C_{\tau,\tau} \int d\textbf{p}' f_{\tau}(\textbf{r},\textbf{p})[\ln(\epsilon(\textbf{p}-\textbf{p}')^{2}+1)]^{2}         \nonumber \\
&&  + \frac{1}{\rho_{0}}C_{\tau,\tau'} \int d\textbf{p}' f_{\tau'}(\textbf{r},\textbf{p})[\ln(\epsilon(\textbf{p}-\textbf{p}')^{2}+1)]^{2}
\nonumber \\
&&  + \alpha\frac{\rho}{\rho_{0}}+\beta\frac{\rho^{\gamma}}{\rho_{0}^{\gamma}}+
E_{sym}^{loc}(\rho)\delta^{2} + \frac{\partial E_{sym}^{loc}(\rho)}{\partial\rho}\rho\delta^{2} \nonumber \\
&&  + E_{sym}^{loc}(\rho)\rho\frac{\partial\delta^{2}}{\partial\rho_{\tau}}  .
\end{eqnarray}
Here $\tau\neq\tau'$, $\partial\delta^{2}/\partial\rho_{n}=4\delta\rho_{p}/\rho^{2}$ and $\partial\delta^{2}/\partial\rho_{p}=-4\delta\rho_{n}/\rho^{2}$. The nucleon effective (Landau) mass in nuclear matter of isospin asymmetry $\delta=(\rho_{n}-\rho_{p})/(\rho_{n}+\rho_{p})$ with $\rho_{n}$ and $\rho_{p}$ being the neutron and proton density, respectively, is calculated through the potential as $m_{\tau}^{\ast}=m_{\tau}/ \left(1+\frac{m_{\tau}}{|\textbf{p}|}|\frac{dU_{\tau}}{d\textbf{p}}|\right)$ with the free mass $m_{\tau}$ at Fermi momentum $\textbf{p}=\textbf{p}_{F}$. Shown in Fig. 2 is the single-particle potentials with the mass splittings of $m_{n}^{\ast}>m_{p}^{\ast}$ and $m_{n}^{\ast}<m_{p}^{\ast}$. It should be noticed that the neutron and proton optical potentials are across with increasing the momentum for the mass splitting of $m_{n}^{\ast}>m_{p}^{\ast}$. The case of $m_{n}^{\ast}<m_{p}^{\ast}$ enlarges the difference in the neutron-rich matter. The variance of optical potential influences the dynamical evolution of nucleons and the kinetic energy spectra of the neutron to proton ratios are changed \cite{Fe12a}. It was found that the neutron/proton ratio at high transverse momenta (kinetic energies) and elliptic flow difference between neutron and proton sensitively are sensitive probes for constraining the isospin splitting. The isospin splitting of nucleon effective mass in neutron-rich nuclear matter has been planning in the RIKEN-SAMURAI experiments.

The dynamics of mesons, hyperons and antiprotons is influenced by the mean-field potentials. The Hamiltonian is constructed as follows
\begin{eqnarray}
H_{\nu}&& = \sum_{i=1}^{N_{\nu}}\left( V_{i}^{\textrm{Coul}} + \omega_{\nu}(\textbf{p}_{i},\rho_{i}) \right).
\end{eqnarray}
Here $\nu$ denotes the hadron species and the Coulomb potential is evaluated by point-like charged particles as
\begin{equation}
V_{i}^{\textrm{Coul}}=\sum_{j=1}^{N_{h}}\frac{e_{i}e_{j}}{r_{ij}}.
\end{equation}
The $N_{\nu}$ and $N_{h}$ are the total numbers of the same species and all charged particles. The in-medium energy $\omega_{\nu}(\textbf{p}_{i},\rho_{i})$ is related to the hadron momentum and baryon density, which is usually evaluated from the scalar and vector potential components to hadron self-energies.

The energy of pion in the nuclear medium is composed of the isoscalar and isovector contributions as follows
\begin{equation}
\omega_{\pi}(\textbf{p}_{i},\rho_{i}) = \omega_{isoscalar}(\textbf{p}_{i},\rho_{i})+C_{\pi}\tau_{z}\delta (\rho/\rho_{0})^{\gamma_{\pi}}.
\end{equation}
The coefficient $C_{\pi}= \rho_{0} \hbar^{3}/(4f^{2}_{\pi}) = 36$ MeV, and the isospin quantities are taken as $\tau_{z}=$ 1, 0, and -1 for $\pi^{-}$, $\pi^{0}$ and $\pi^{+}$, respectively. The isospin asymmetry $\delta=(\rho_{n}-\rho_{p})/(\rho_{n}+\rho_{p})$ and the quantity $\gamma_{\pi}$ adjusts the isospin splitting of pion optical potential. Usually, we take the $\gamma_{\pi} = 2$ having the obvious difference of $\pi^{-}$-nucleon and $\pi^{+}$-nucleon potentials in the dense neutron-rich matter. The isoscalar pion energy in the nuclear medium is calculated with the $\Delta$-hole model \cite{Br75,Fe15a,Fe17}. The in-medium dispersion relation consists of a pion branch (smaller value) and a $\Delta$-hole (larger value) branch. The $\pi$-like energy in nuclear medium decreases with increasing the baryon density. However, the $\Delta$-like energy is just opposite. The dispersion relation reads
\begin{eqnarray}
\omega_{isoscalar}(\textbf{p}_{i},\rho_{i}) = &&  S_{\pi}(\textbf{p}_{i},\rho_{i}) \omega_{\pi-like}(\textbf{p}_{i},\rho_{i}) +    \nonumber \\
&&        S_{\Delta}(\textbf{p}_{i},\rho_{i}) \omega_{\Delta-like}(\textbf{p}_{i},\rho_{i}) .
\end{eqnarray}
The probabilities of the pion component satisfy the relation
\begin{equation}
S_{\pi}(\textbf{p}_{i},\rho_{i}) + S_{\Delta}(\textbf{p}_{i},\rho_{i}) = 1
\end{equation}
The value of the probability is determined from the pion self-energy as \cite{Xi93}
\begin{equation}
S(\textbf{p}_{i},\rho_{i}) = \frac{1}{1-\partial \Pi (\omega)/\partial\omega^{2}},
\end{equation}
where the pion self-energy is given by
\begin{equation}
\Pi = \textbf{p}_{i}^{2}\frac{\chi}{1 - g\prime\chi},
\end{equation}
with the Migdal parameter $g\prime\sim$0.6 and
\begin{equation}
\chi = -\frac{8}{9}\left(\frac{f_{\Delta}}{m_{\pi}}\right)^{2} \frac{\omega_{\Delta}\rho\hbar^{3}}{\omega_{\Delta}^{2}-\omega^{2}} \exp(-2\textbf{p}_{i}^{2}/b^{2}).
\end{equation}
The $\omega_{\Delta}=\sqrt{m_{\Delta}^{2}+\textbf{p}_{i}^{2}}-m_{N}$, the $m_{\pi}$, $m_{N}$ and $m_{\Delta}$ are the masses of pion, nucleon and delta, respectively. The $\pi N\Delta$ coupling constant $f_{\Delta}\sim 2 $ and the cutoff factor $b\sim 7 m_{\pi}$. Two eigenvalues of $\omega_{\pi-like}$ and $\omega_{\Delta-like}$ are obtained from the pion dispersion relation as
\begin{equation}
\omega^{2} = \textbf{p}_{i}^{2} + m_{\pi}^{2} + \Pi(\omega).
\end{equation}

\begin{figure*}
\includegraphics[width=16 cm]{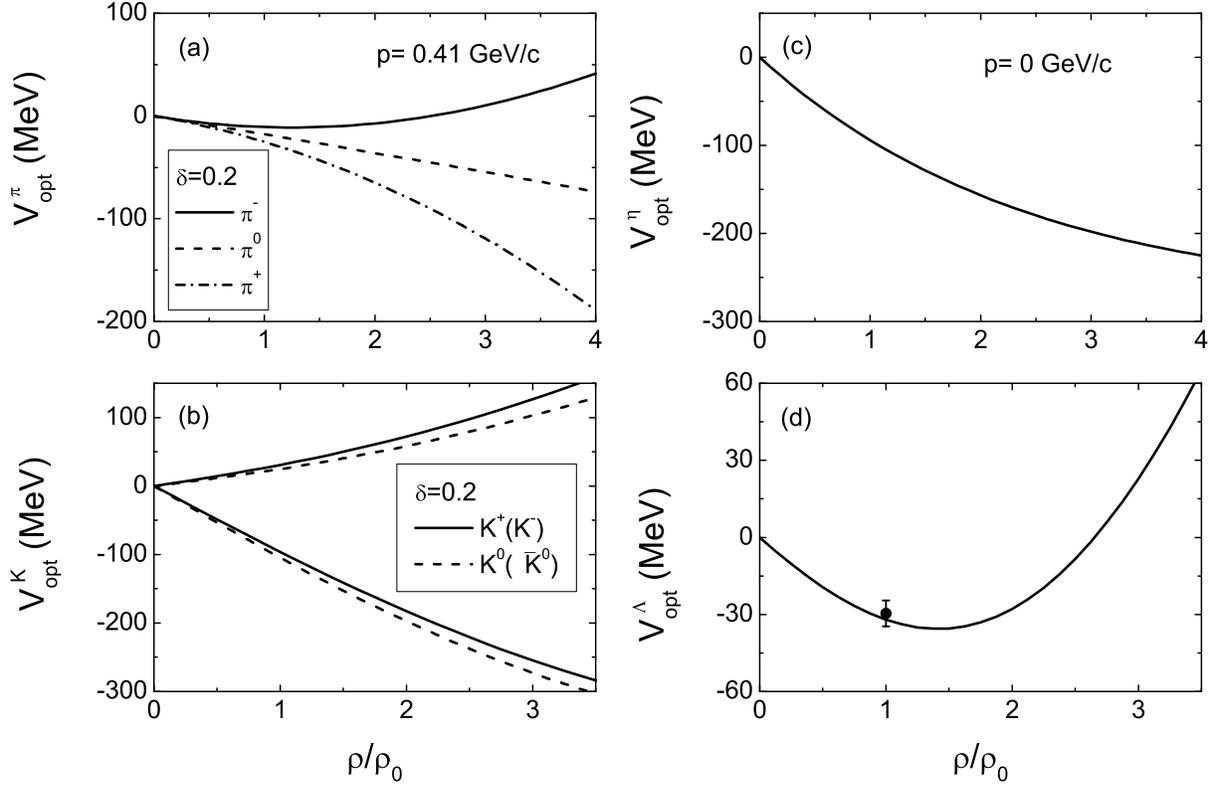}
\caption{Density dependence of the optical potentials for pions, etas, kaons and hyperons in dense nuclear matter. The empirical value of $\Lambda$ potential extracted from hypernucleus experiments is denoted by the solid circle \cite{Mi88}.}
\end{figure*}

The Hamiltonian of $\eta$ is composed of
\begin{equation}
H_{\eta} = \sum_{i=1}^{N_{\eta}}\left( \sqrt{m_{\eta}^{2}+\textbf{p}_{i}^{2}} + V_{\eta}^{opt}(\textbf{p}_{i},\rho_{i}) \right).
\end{equation}
The eta optical potential is evaluated from the dispersion relation based on the chiral perturbation theory \cite{Ni08} as
\begin{equation}
\omega_{\eta}(\textbf{p}_{i},\rho_{i}) = \sqrt{\left(m^{2}_{\eta} - a_{\eta}\rho_{s}\right) \left(1+b_{\eta}\rho_{s}\right)^{-1} + \textbf{p}_{i}^{2}}
\end{equation}
with $a_{\eta}=\hbar^{3}\frac{\Sigma_{\eta N}}{f^{2}_{\pi}}$ and $b_{\eta}=\hbar^{2}\frac{\kappa}{f^{2}_{\pi}}$. The pion decay constant $f_{\pi}=$92.4 MeV, $\Sigma_{\eta N}=$280 MeV and $\kappa =$0.4 fm. The optical potential is given by $V_{\eta}^{opt}(\textbf{p}_{i},\rho_{i}) = \omega_{\eta}(\textbf{p}_{i},\rho_{i}) - \sqrt{m_{\eta}^{2}+\textbf{p}_{i}^{2}}$ with the eta mass $m_{\eta}$=547 MeV. The value of $V_{\eta}^{opt}=$ -94 MeV is obtained with zero momentum and saturation density $\rho=\rho_{0}$.

The energies of kaon and anti-kaon in nuclear medium are calculated with the chiral Lagrangian approach as follows \cite{Fe15a,Sc97}
\begin{eqnarray}
\omega_{K}(\textbf{p}_{i},\rho_{i})= && [m_{K}^{2}+\textbf{p}_{i}^{2}-a_{K}\rho_{i}^{S}
-\tau_{3}c_{K}\rho_{i3}^{S}                              \nonumber \\
&& +(b_{K}\rho_{i}+\tau_{3}d_{K}\rho_{i3})^{2}]^{1/2}+b_{K}\rho_{i}              \nonumber \\
&& +\tau_{3}d_{K}\rho_{i3}
\end{eqnarray}
and
\begin{eqnarray}
\omega_{\overline{K}}(\textbf{p}_{i},\rho_{i})= && [m_{\overline{K}}^{2}+\textbf{p}_{i}^{2}-a_{\overline{K}}\rho_{i}^{S}
-\tau_{3}c_{K}\rho_{i3}^{S}                              \nonumber \\
&& +(b_{K}\rho_{i}+\tau_{3}d_{K}\rho_{i3})^{2}]^{1/2}-b_{K}\rho_{i}                   \nonumber \\
&& -\tau_{3}d_{K}\rho_{i3},
\end{eqnarray}
respectively. Here the $b_{K}=3\hbar^{3}/(8f_{\pi}^{2})\approx$ 0.337 GeVfm$^{3}$, the $a_{K}$ and $a_{\overline{K}}$ are 0.18 GeV$^{2}$fm$^{3}$ and 0.31 GeV$^{2}$fm$^{3}$. The $\tau_{3}$=1 and -1 for the isospin pair K$^{+}$($\overline{K}^{0}$) and K$^{0}$(K$^{-}$), respectively. The parameters $c_{K}$=0.0298 GeV$^{2}$fm$^{3}$ and $d_{K}$=0.111 GeVfm$^{3}$ determine the isospin splitting of kaons in neutron-rich nuclear matter. The optical potential of kaon is derived from the in-medium energy as $V_{opt}(\textbf{p},\rho)=\omega(\textbf{p},\rho)-\sqrt{\textbf{p}^{2}+m_{K}^{2}}$. The optical potentials being the values of 31 MeV, 25 MeV, -96 MeV and 104 MeV at saturation density with the isospin asymmetry of 0.2 are obtained for $K^{+}$, $K^{0}$, $K^{-}$ and $\overline{K}^{0}$ in nuclear medium, respectively. The values of $m^{\ast}_{K}/m_{K}$=1.056 and $m^{\ast}_{\overline{K}}/m_{\overline{K}}$= 0.797 at normal baryon density are concluded with the parameters in isospin symmetric nuclear matter. The effective mass is used to evaluate the threshold energy for kaon and antikaon production, e.g., the threshold energy in the pion-baryon collisions $\sqrt{s_{th}}=m^{\ast}_{Y} + m^{\ast}_{K}$.

The in-medium dispersion relation for hyperons reads as
\begin{equation}
\omega_{Y}(\textbf{p}_{i},\rho_{i})=\sqrt{(m_{Y}+\Sigma_{S}^{Y})^{2}+\textbf{p}_{i}^{2}} + \Sigma_{V}^{Y},
\end{equation}
The hyperon self-energies are evaluated on the basis of the light-quark counting rules, i.e., $\Lambda$ and $\Sigma$ being assumed to be two thirds of nucleon self-energies, the $\Xi$ self-energy being one third of nucleon's ones. Namely, for hyperons $\Sigma_{S}^{\Lambda,\Sigma}= 2 \Sigma_{S}^{N}/3$, $\Sigma_{V}^{\Lambda,\Sigma}= 2 \Sigma_{V}^{N}/3$, $\Sigma_{S}^{\Xi}= \Sigma_{S}^{N}/3$ and $\Sigma_{V}^{\Xi}= \Sigma_{V}^{N}/3$. The antibaryon energy is computed from the G-parity transformation of baryon potential as
\begin{equation}
\omega_{\overline{B}}(\textbf{p}_{i},\rho_{i})=\sqrt{(m_{\overline{B}}+\Sigma_{S}^{\overline{B}})^{2}+\textbf{p}_{i}^{2}} + \Sigma_{V}^{\overline{B}}
\end{equation}
with $\Sigma_{S}^{\overline{B}}=\Sigma_{S}^{B}$ and $\Sigma_{V}^{\overline{B}}=-\Sigma_{V}^{B}$.
The nuclear scalar $\Sigma_{S}^{N}$ and vector $\Sigma_{V}^{N}$ self-energies are computed from the well-known relativistic mean-field model with the NL3 parameter ($g_{\sigma N}^{2}$=80.8, $g_{\omega N}^{2}$=155 and $g_{\rho N}^{2}$=20). The optical potential of baryon or antibaryon is derived from the in-medium energy as $V_{opt}(\textbf{p},\rho)=\omega(\textbf{p},\rho)-\sqrt{\textbf{p}^{2}+m^{2}}$. The relativistic self-energies are used for the construction of hyperon and antibaryon potentials only. A very deep antiproton-nucleus potential is obtained with the G-parity approach being $V_{opt}(\textbf{p}=0,\rho=\rho_{0}) = -655 $ MeV. From fitting the antiproton-nucleus scattering \cite{La09} and the real part of phenomenological antinucleon-nucleon optical potential \cite{Co82}, a factor $\xi$ is introduced in order to moderately evaluate the optical potential as $\Sigma_{S}^{\overline{N}}=\xi\Sigma_{S}^{N}$ and $\Sigma_{V}^{\overline{N}}=-\xi\Sigma_{V}^{N}$ with $\xi$=0.25, which leads to the strength of $V_{\overline{N}}=-164$ MeV at the normal nuclear density $\rho_{0}$=0.16 fm$^{-3}$. It should be noted that the scaling approach violates the fundamental G-symmetry. The antihyperon potentials exhibit strongly attractive interaction in nuclear medium, e.g., the strengths at saturation density being -436 MeV and -218 MeV for $\overline{\Lambda}$ and $\overline{\Xi}$, respectively. The optical potentials will affect the dynamics of hyperons, consequently the production of hypernuclei. Shown in Fig. 3 is the optical potentials for pions, etas, kaons and hyperons. The empirical value extracted from hypernucleus experiments \cite{Gi95,Mi88} can be well reproduced by the optical potential. The same approach is used for the $\Sigma$ propagation in nuclear medium. The optical potentials are of importance in binding hyperon into a nucleus.

\begin{figure*}
\includegraphics[width=16 cm]{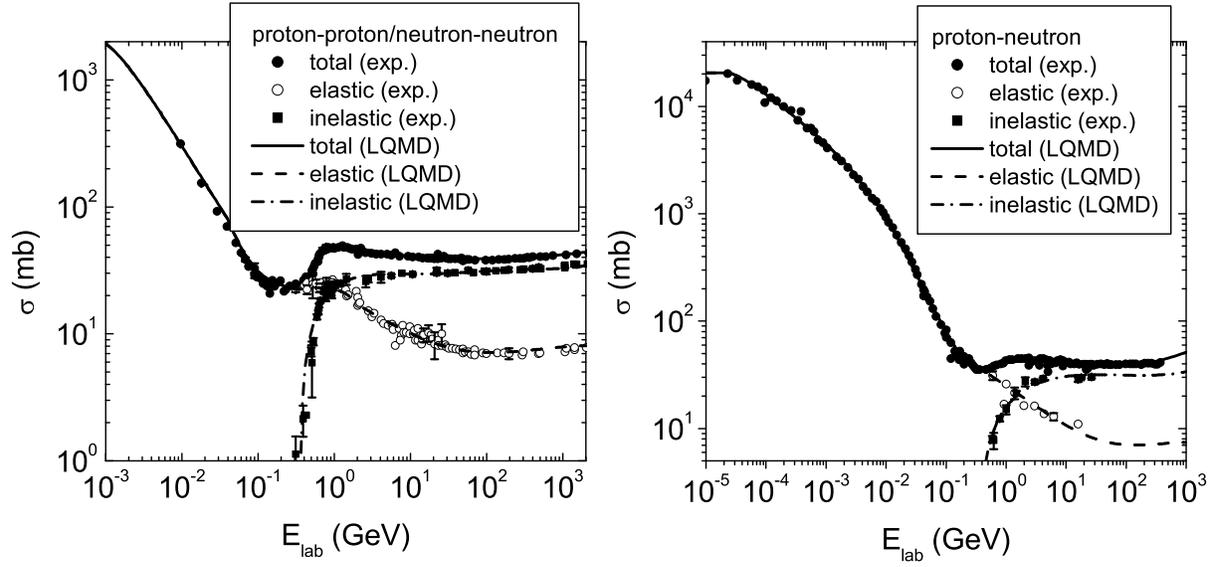}
\caption{Nucleon-nucleon total, elastic and inelastic cross sections in free space parameterized in the LQMD model as a function of nucleon incident energy in the laboratory frame.}
\end{figure*}

\begin{figure*}
\includegraphics[width=16 cm]{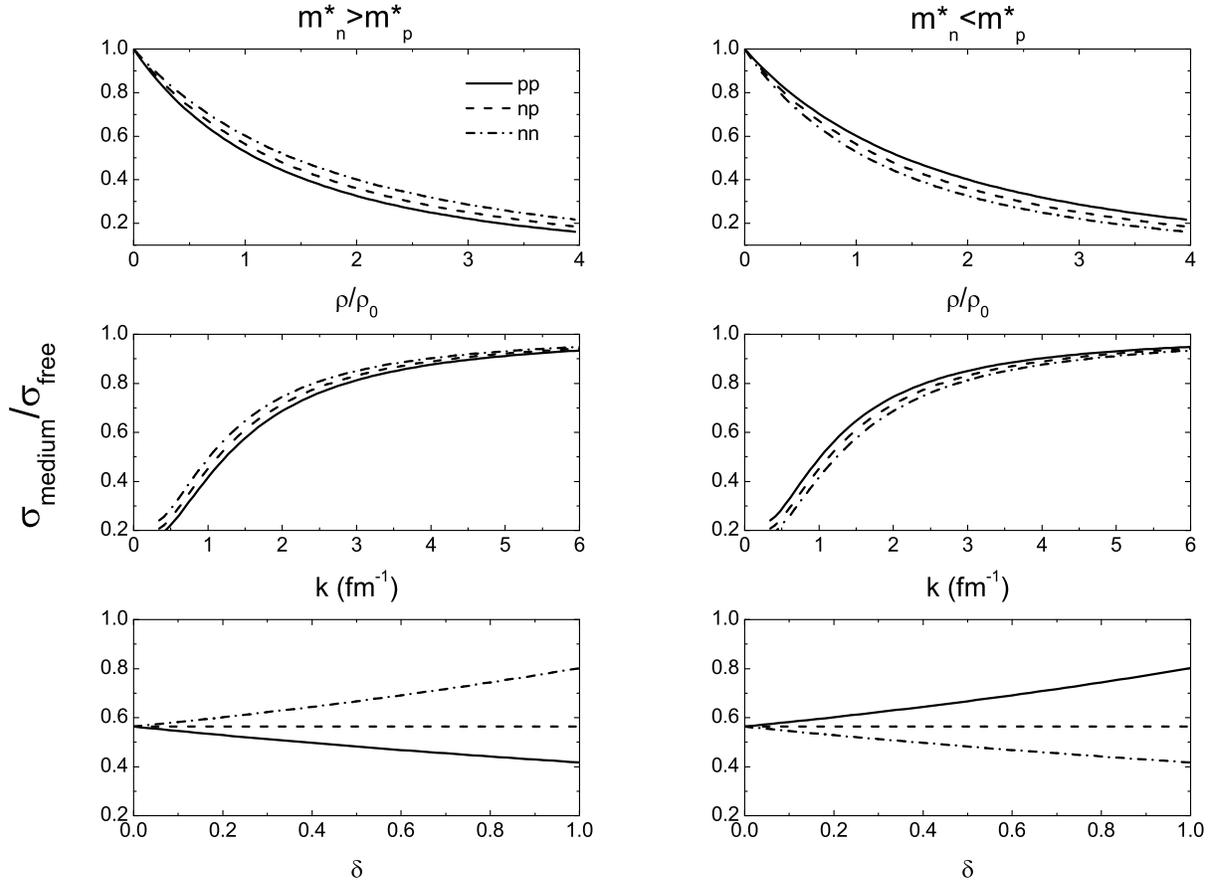}
\caption{In-medium nucleon-nucleon elastic cross section scaled by nucleon effective mass as functions of baryon density ($\delta=0.2$), nucleon momentum ($\delta=0.2$) and isospin asymmetry ($\rho=\rho_{0}$).}
\end{figure*}

The probability of two-particle or three-particle collision is sampled with a Monte Carlo procedure, in which the scattering of two particles is determined by a geometrical minimum distance criterion $d\leq\sqrt{0.1\sigma_{tot}/\pi}$ fm weighted by the Pauli blocking of the final states \cite{Be88,Ai87}. Here, the total NN cross section $\sigma_{tot}$ in mb is the sum of the elastic and all inelastic baryon-baryon collisions. The channel probability is evaluated by comparing the channel to the total cross section as $P_{ch}=\sigma_{ch}/\sigma_{tot}$. The choice of reaction channel is randomly performed by the weight of the channel probability. The total, elastic and inelastic NN cross sections are parameterized in accordance with the available experimental data \cite{Ca93} as shown in Fig. 4. One expects the in-medium NN elastic collisions are reduced in comparison with the free-space ones. The in-medium elastic cross section is scaled according to the effective mass that was used in the BUU model \cite{De02}. The in-medium elastic cross section is scaled according to the effective mass through $\sigma_{NN}^{medium}=(\mu^{\ast}_{NN}/\mu_{NN})^{2}\sigma_{NN}^{free}$ with the $\mu^{\ast}_{NN}$ and $\mu_{NN}$ being the reduced masses of colliding nucleon pairs in the medium and in the free space \cite{Li05b}, respectively. Shown in Fig. 5 is a comparison of the scaling factor as functions of baryon density, nucleon momentum and isospin asymmetry for the different mass splittings of $m_{n}^{\ast}>m_{p}^{\ast}$ and $m_{n}^{\ast}<m_{p}^{\ast}$, respectively. It is interesting to notice that a splitting of $nn$ (neutron-neutron), $pn$ (proton-neutron) and $pp$ (proton-proton) exists because of the mixing of neutron and proton effective masses. The mass splitting results in an opposite splitting of the in-medium cross sections of $nn$ and $pp$ channels. The NN elastic cross section impacts the collective flows in heavy-ion collisions, in particular the balance energy \cite{Fe12b}. It is found that the in-medium elastic cross sections play a significant role in isospin emission and result in a flatter distribution for transverse flows and elliptic flows of free nucleons compared with the in-vacuum ones. However, the flow difference of neutrons and protons is independent on the in-medium modifications of NN cross sections and more sensitive to the stiffness of symmetry energy.

\begin{figure*}
\includegraphics[width=16 cm]{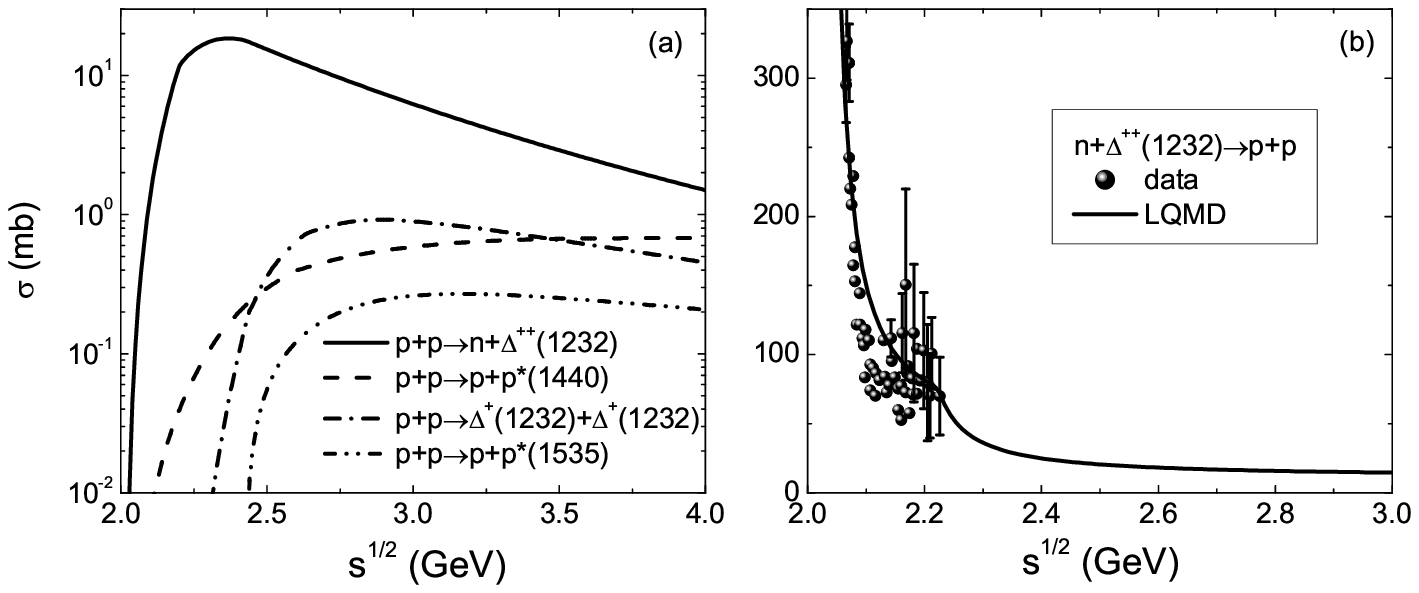}
\caption{Production and inverse absorption cross sections of the resonances $\Delta$(1232), $N^{\ast}$(1440) and $N^{\ast}$(1535). The experimental data of the channel N$\Delta\rightarrow$NN are shown for comparison \cite{Ho96a,Ho96b}.}
\end{figure*}

\begin{figure*}
\includegraphics[width=16 cm]{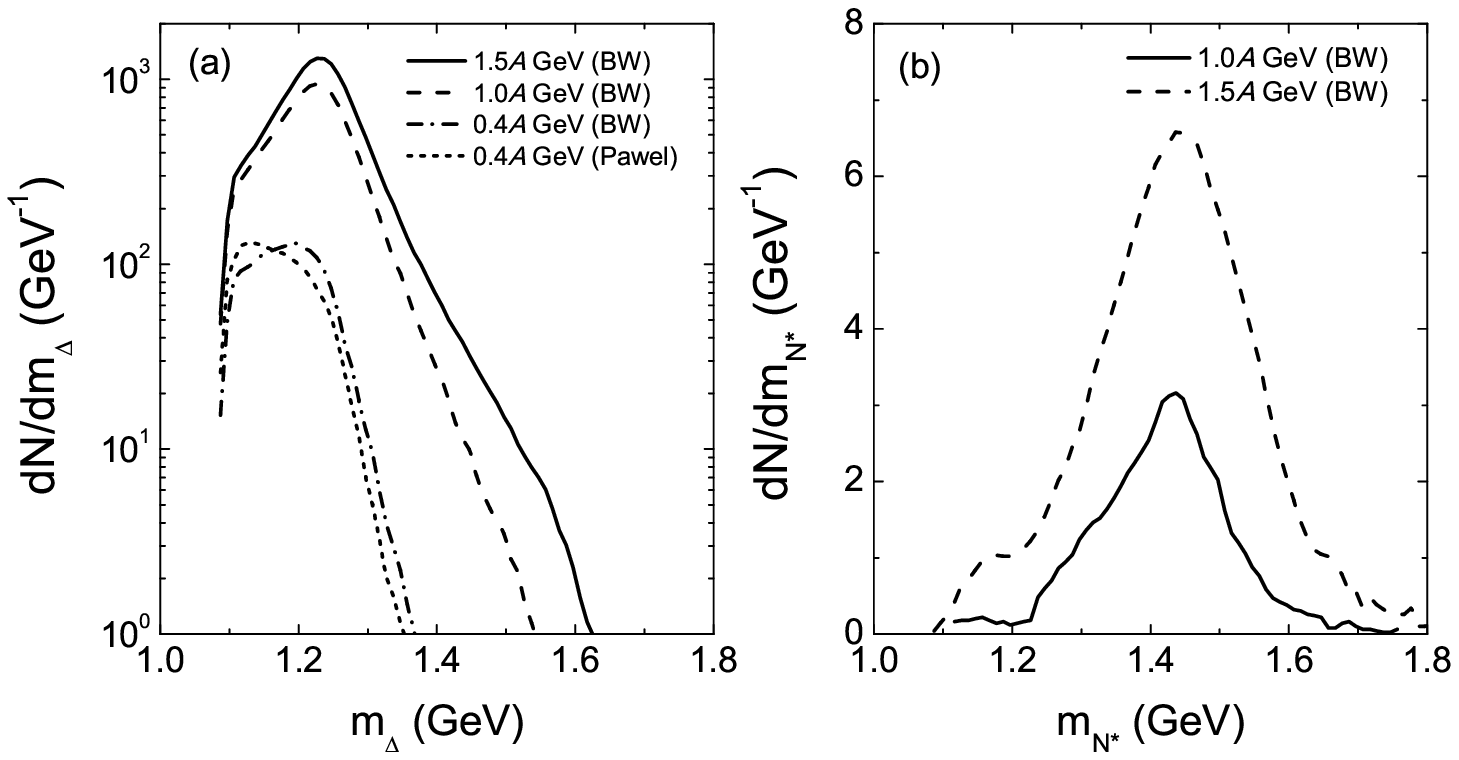}
\caption{Mass distributions of the resonances $\Delta$(1232) and $N^{\ast}$(1440) in the $^{197}$Au+$^{197}$Au reaction at the incident energies of 400, 1000 and 1500 \emph{A} MeV, respectively.}
\end{figure*}

The inelastic NN collisions are complicated and a number of reaction channels are opened and coupled to each other.
The primary products in NN collisions are the resonances and their decay processes, such as $\Delta$(1232), $N^{\ast}$(1440), $N^{\ast}$(1535) etc. We have included the reaction channels as follows:
\begin{eqnarray}
&& NN \leftrightarrow N\triangle, \quad  NN \leftrightarrow NN^{\ast}, \quad  NN
\leftrightarrow \triangle\triangle,  \nonumber \\
&& \Delta \leftrightarrow N\pi,  N^{\ast} \leftrightarrow N\pi,  NN \leftrightarrow NN\pi (s-state),  \nonumber \\
&& N^{\ast}(1535) \leftrightarrow N\eta.
\end{eqnarray}
Here hadron-hadron collisions take place as two-body process and three-body ($s-$state pion production) reactions.
The momentum-dependent decay widths are used for the resonances of $\Delta$(1232) and $N^{\ast}$(1440) \cite{Fe09,Hu94}. We have taken a constant width of $\Gamma$=150 MeV for the $N^{\ast}$(1535) decay. Elastic scattering of nucleon-resonance ($NR\rightarrow NR$) and resonance-resonance ($RR\rightarrow RR$) collisions and inelastic collisions of nucleon-resonance ($NR\rightarrow NN$, $NR\rightarrow NR\prime$) and resonance-resonance ($RR\rightarrow NN$, $RR\rightarrow RR\prime$, $R$ and $R\prime$ being different resonances), have been included in the model. A parameterized cross section is used in the LQMD model for the channel of N$\Delta\rightarrow$NN by fitting the available experimental data \cite{Ho96a,Ho96b} similar to the GiBUU model \cite{Wo93}. Calculations based on relativistic Dirac-Brueckner also favor the decrease trend of the $\Delta$ absorption cross section with increasing energy in nuclear medium at low densities (less than 2$\rho_{0}$) \cite{Ha87}. We used the parametrized cross sections calculated by the one-boson exchange model \cite{Hu94} for resonance production and the absorption of $N^{\ast}$ with the detailed balancing principle. Shown in Fig. 6 is the energy dependence of resonance production in proton-proton collisions. It is obvious that the $\Delta$(1232) production is dominant in the medium-energy heavy-ion collisions. The mass of resonance is usually sampled with the Breit-Wigner (BW) distribution as
\begin{equation}
p(m_{r})=\frac{1}{1+4[(m_{r}-m_{r}^{0})/\Gamma]^{2}}
\end{equation}
with the peak mass $m_{r}^{0}$ and the decay width $\Gamma$. Shown in Fig. 7 is the mass distributions of the resonances $\Delta$(1232) and $N^{\ast}$(1440) in the $^{197}$Au+$^{197}$Au collisions. A modified formula proposed by Pawel \cite{Da91} is also shown for the $\Delta$(1232) mass sampling at the incident energy of 0.4 GeV/nucleon. The distribution is slightly moved towards the lower mass region with the Pawel's formula in comparison with the standard BW form.

\begin{figure*}
\includegraphics[width=18 cm]{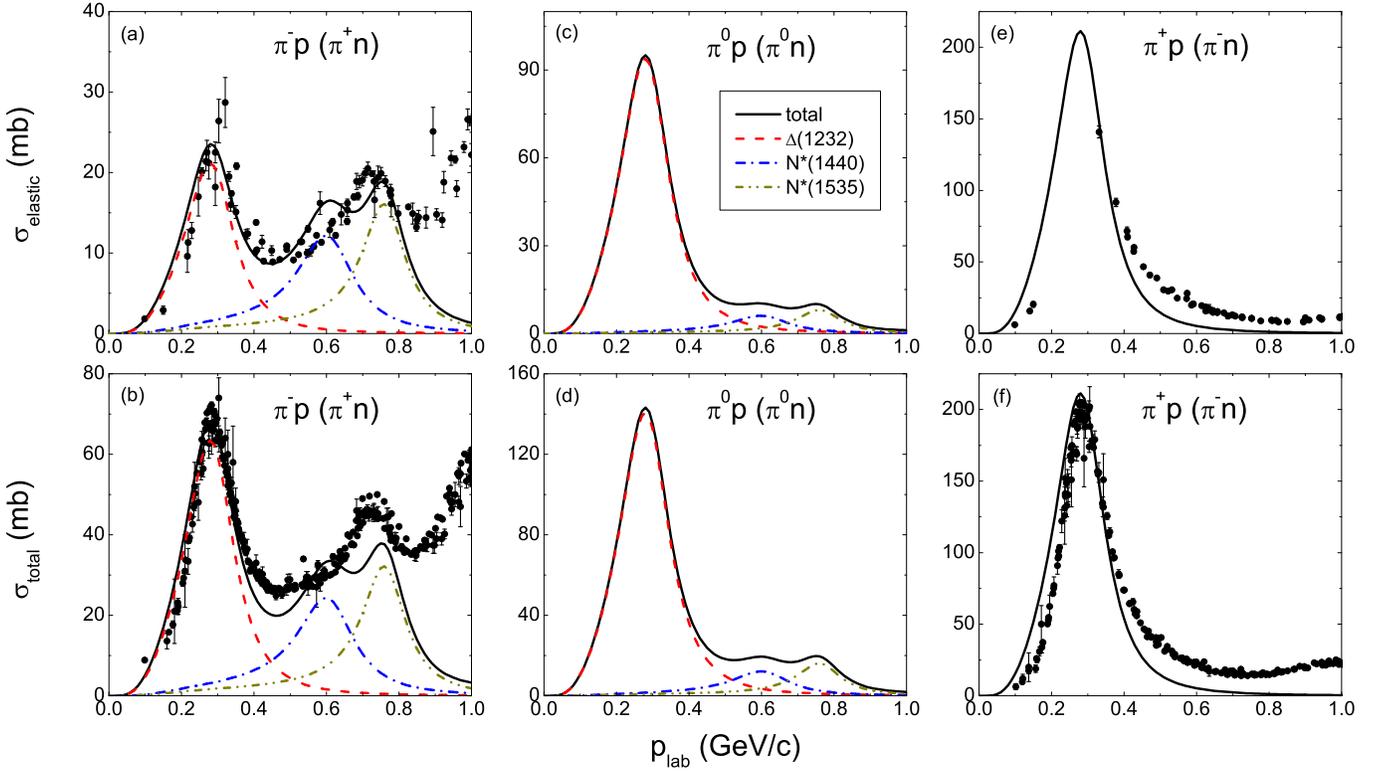}
\caption{(Color online) The elastic (upper panel) and total (down panel) pion-nucleon scattering cross sections contributed from different resonances. The available data are taken from the PDG collaboration \cite{Ol14}. }
\end{figure*}

The cross section of pion-nucleon scattering is evaluated with the Breit-Wigner formula as the form of
\begin{eqnarray}
\sigma_{\pi N\rightarrow R}(\sqrt{s}) = \sigma_{max}(|\textbf{p}_{0}/\textbf{p}|)^{2}\frac{0.25\Gamma^{2}(\textbf{p})}
{0.25\Gamma^{2}(\textbf{p})+(\sqrt{s}-m_{0})^{2}},
\end{eqnarray}
where the $\textbf{p}$ and $\textbf{p}_{0}$ are the momenta of pions at the energies of $\sqrt{s}$ and $m_{0}$, respectively, and $m_{0}$ being the centroid of resonance mass, e.g., 1.232 GeV, 1.44 GeV and 1.535 GeV for $\Delta$(1232), $N^{\ast}$(1440), and $N^{\ast}$(1535), respectively. The maximum cross section $\sigma_{max}$ is taken from fitting the total cross sections of the available experimental data. For example, 200 mb, 133.3 mb, and 66.7 mb for $\pi^{+}+p\rightarrow \Delta^{++}$ ($\pi^{-}+n\rightarrow \Delta^{-}$), $\pi^{0}+p\rightarrow \Delta^{+}$ ($\pi^{0}+n\rightarrow \Delta^{0}$) and $\pi^{-}+p\rightarrow \Delta^{0}$ ($\pi^{+}+n\rightarrow \Delta^{+}$), respectively. And 24 mb, 12 mb, 32 mb, 16 mb for $\pi^{-}+p\rightarrow N^{\ast 0}(1440)$ ($\pi^{+}+n\rightarrow N^{\ast +}(1440)$), $\pi^{0}+p\rightarrow N^{\ast +}(1440)$ ($\pi^{0}+n\rightarrow N^{\ast 0}(1440)$), $\pi^{-}+p\rightarrow N^{\ast 0}(1535)$ ($\pi^{+}+n\rightarrow N^{\ast +}(1535)$) and $\pi^{0}+p\rightarrow N^{\ast +}(1535)$ ($\pi^{0}+n\rightarrow N^{\ast 0}(1535)$), respectively. Shown in Fig. 8 is a comparison of the elastic and total cross sections in the pion-nucleon collisions with the available data from the PDG collaboration \cite{Ol14}. The total cross sections include the sum of three resonances and the contributions of strangeness production. It is obvious that the more resonances are needed to be implemented at the pion momentum above 0.5 GeV/c. The spectra can be reproduced nicely well with the resonance approach, in particular in the domain of $\Delta$ resonance momenta (0.298 GeV/c). The pion-nucleon scattering dominates the energy deposition in the pion-induced reactions, which contributes the fragmentation of target nuclei \cite{Fe16a}.

\begin{figure*}
\includegraphics[width=16 cm]{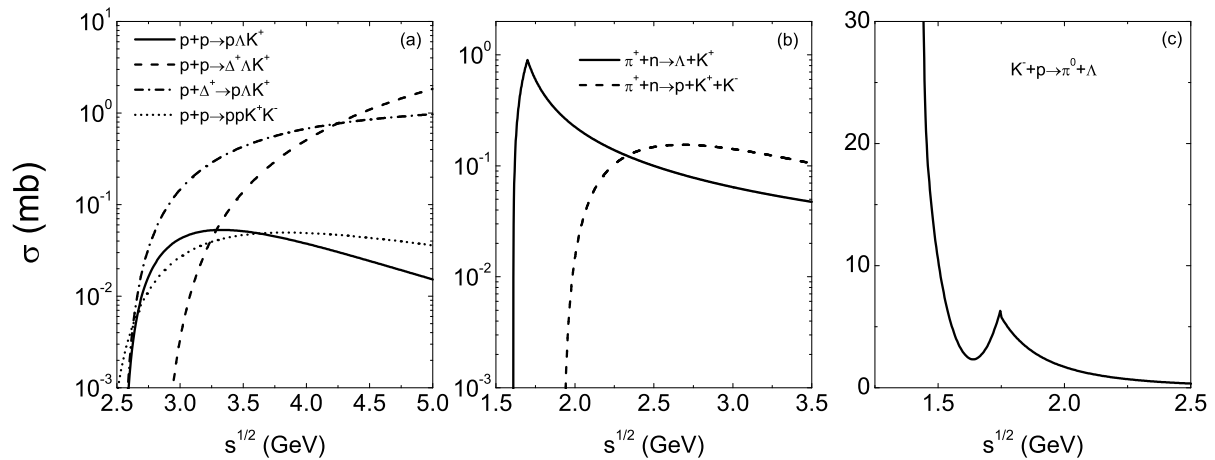}
\caption{Elementary cross sections in the production of strange particles.}
\end{figure*}

The strangeness and vector mesons ($\rho$, $\omega$) are created in inelastic hadron-hadron collisions without intermediate resonances. There are the 2-body and 3-body reaction channels as follows:
\begin{eqnarray}
&& BB \rightarrow BYK,  BB \rightarrow BBK\overline{K},  B\pi(\eta) \rightarrow YK,  YK \rightarrow B\pi,     \nonumber \\
&& B\pi \rightarrow NK\overline{K}, Y\pi \rightarrow B\overline{K}, \quad  B\overline{K} \rightarrow Y\pi, \quad YN \rightarrow \overline{K}NN,  \nonumber \\
&& NN \rightarrow NN\rho, NN \rightarrow NN\omega.
\end{eqnarray}
Here the B stands for (N, $\triangle$, N$^{\ast}$) and Y($\Lambda$, $\Sigma$), K(K$^{0}$, K$^{+}$) and $\overline{K}$($\overline{K}^{0}$, K$^{-}$). The parameterized cross sections of each isospin channel $BB \rightarrow BYK$ \cite{Ts99} are used in the calculation. We take the parametrizations of the channels $B\pi \rightarrow YK$ \cite{Ts94,Ts95} besides the $N\pi \rightarrow \Lambda K$ reaction \cite{Cu84}. The results are close to the experimental data at near threshold energies. The cross section of antikaon production in inelastic hadron-hadron collisions is taken as the same form of the parametrization used in the hadron string dynamics (HSD) calculations \cite{Ca97}. Furthermore, the elastic scattering and strangeness-exchange reaction between strangeness and baryons have been considered through the channels of $KB \rightarrow KB$, $YB \rightarrow YB$ and $\overline{K}B \rightarrow \overline{K}B$ and we use the parametrizations in Ref. \cite{Cu90}. The charge-exchange reactions between the $KN \rightarrow KN$ and $YN \rightarrow YN$ channels are included by using the same cross sections with the elastic scattering, such as $K^{0}p\rightarrow K^{+}n$, $K^{+}n\rightarrow K^{0}p$ etc \cite{Fe13b}. The cross sections for $\rho$ and $\omega$ production are taken from the fitting of experimental data \cite{Li01}. Shown in Fig. 9 is the elementary cross section of different channel associated with the strangeness reactions. It is obvious that the $\Delta$-nucleon collisions are dominant in the production of strange particles. The large absorption cross section in the K$^{-}$-nucleon collisions retards the K$^{-}$ emission in near-threshold energy heavy-ion collisions, but enhances the hyperon production. The pion induced reactions have considerable contributions on the strangeness production, roughly, one third to one fourth K$^{+}$ yields being from the pion channels in heavy-ion collisions around the threshold energies \cite{Fe13a}.

In order to describe the antiproton-nucleus collisions we have further included the annihilation channels, charge-exchange reaction (CEX), elastic (EL) and inelastic scattering as follows \cite{Fe14a,Fe16c}:
\begin{eqnarray}
&& \overline{B}B \rightarrow \texttt{annihilation}(\pi,\eta,\rho,\omega,K,\overline{K},\eta\prime,K^{\ast},\overline{K}^{\ast},\phi),
\nonumber \\
&&  \overline{B}B \rightarrow \overline{B}B (\texttt{CEX, EL}),   \overline{N}N \leftrightarrow \overline{N}\Delta(\overline{\Delta}N), \overline{B}B \rightarrow \overline{Y}Y.
\end{eqnarray}
Here the B strands for nucleon and $\Delta$(1232), Y($\Lambda$, $\Sigma$, $\Xi$), K(K$^{0}$, K$^{+}$) and $\overline{K}$($\overline{K^{0}}$, K$^{-}$). The overline of B (Y) means its antiparticle. The cross sections of these channels are based on the parametrization or extrapolation from available experimental data \cite{Bu12}. The annihilation dynamics in antibaryon-baryon collisions is described by a statistical model with SU(3) symmetry inclusion of all pseudoscalar and vector mesons \cite{Go92}. Pions are the dominant products in the antiproton annihilation on a nucleus. The energy released in the antiproton-nucleon annihilation is mainly deposited in the target nucleus via the pion-nucleon collisions. Hyperons are mainly produced via strangeness exchange reactions in collisions of antikaons and nucleons, which have smaller relative momentum and could be easily captured by the residue nuclei to form hypernuclei. The nuclear fragmentation, hypernuclide formation, particle production, unexpected neutron/proton ratio and isospin effect in antiproton induced reactions have been thoroughly investigated within the LQMD model \cite{Fe16c,Fe16d,Fe17b}.

\section{Nuclear dynamics and isospin effect in heavy-ion collisions}

The LQMD transport model has been developed for many issues in nuclear dynamics, i.e., the neck fragmentation and isospin diffusion in quasi-fission dynamics from Fermi-energy heavy-ion collisions, high-density symmetry energy, in-medium properties of hadrons etc. Motivation of the future experiments is discussed on the topics of hypernuclide production, strangeness dynamics, isospin physics and nuclear phase diagram in heavy-ion collisions and hadron induced reactions. Parts of the results are presented in the article.

\subsection{Neck fragmentation in isotopic nuclear reactions}

Heavy-ion collisions at the Fermi energies (10-100\emph{A} MeV) attract much attention on several topical issues in nuclear physics, i.e., spinodal multifragmentation, liquid-gas phase transition, properties of highly excited nuclei, symmetry energy in the domain of subnormal densities etc \cite{Ma99,Ch04,Co93,Co94,Po95,Wu98,Wu02,De17}. The isospin dynamics in the Fermi-energy heavy-ion collisions is related to several interesting topics, e.g., in-medium nucleon-nucleon cross sections, cluster formation and correlation, isotopic fragment distribution, the symmetry energy in dilute nuclear matter etc. Different mechanisms can coexist and are correlated in heavy-ion collisions at the Fermi energies, such as projectile or target fragmentation, neck emission, preequilibrium emission of light clusters (complex particles), fission of heavy fragments, multifragmentation etc, in which the isospin dependent nucleon-nucleon potential dominates the dynamical processes. The time scales from dynamical and pre-equilibrium emissions to statistical decay of excited systems at equilibrium and the isospin effect in neck fragmentation were investigated in experimentally \cite{Fi14,Ru15}. The LQMD transport model is used to describe the nuclear dynamics in Fermi-energy heavy-ion collisions.

\begin{figure*}
\includegraphics[width=16 cm]{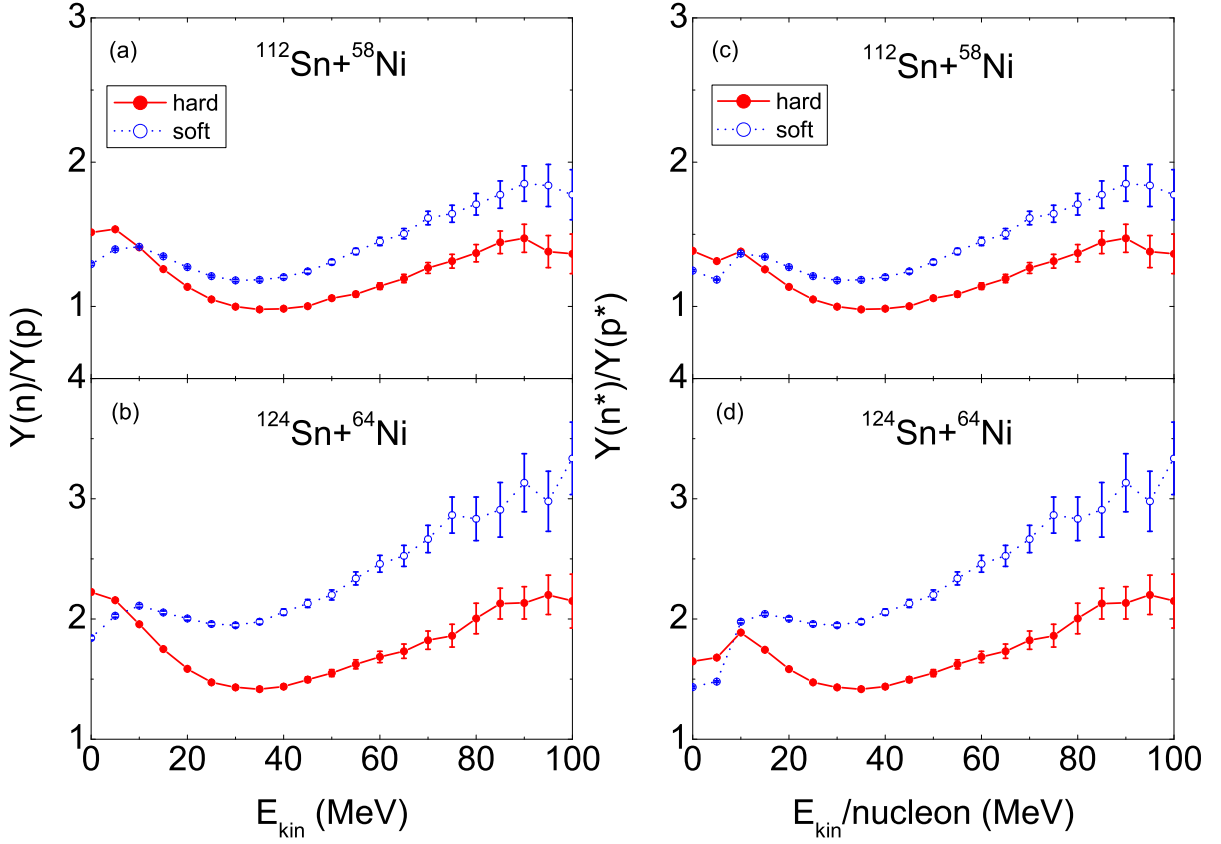}
\caption{(Color online) Kinetic energy spectra of neutron/proton ratios from the yields of free nucleons ((a) and (b)) and 'gas-phase' nucleons (nucleons, hydrogen and helium isotopes) ((c) and (d)) from neck fragmentations in the $^{112}$Sn+$^{58}$Ni and $^{124}$Sn+$^{64}$Ni reactions at the fermi energy of 35 MeV/nucleon within the collision centralities of 6-8 fm.}
\end{figure*}

The composite system formed in Fermi energy heavy-ion collisions is highly excited with an excitation energy up to several tens of MeV/nucleon. The heated system is unstable and fragmentation even multifragmentation takes place. The fragmentation dynamics in the Fermi-energy heavy-ion collisions is described by the LQMD model. The primary fragments are constructed in phase space with a coalescence model, in which nucleons at freeze-out are considered to belong to one cluster with the relative momentum smaller than $P_{0}$ and with the relative distance smaller than $R_{0}$ (here $P_{0}$ = 200 MeV/c and $R_{0}$ = 3 fm). At the freeze-out, the primary fragments are highly excited. The de-excitation of the fragments is described within the GEMINI code \cite{Ch88}. Particles produced from the neck fragmentations in heavy-ion collisions could be probes of the low-density phase diagram, which are constrained within the midrapidities ($|y/y_{proj}|<$0.3) in semicentral nuclear collisions. Shown in Fig. 10 is the kinetic energy spectra of neutron/proton ratios from the yields of free nucleons and 'gas-phase' nucleons (nucleons, hydrogen and helium isotopes) from the neck fragmentations in the $^{112}$Sn+$^{58}$Ni and $^{124}$Sn+$^{64}$Ni reactions at a beam energy of 35 MeV/nucleon with the different symmetry energies \cite{Fe16b}. A larger value of the n/p ratio with softening symmetry energy is found, in particular at the kinetic energies above the Fermi energy (36 MeV), which is caused from the fact that more repulsive force enforced on neutrons in dilute nuclear matter. Calculations are performed without the momentum dependent interaction and the set of parameter of PAR2 is taken.

\begin{figure*}
\includegraphics[width=16 cm]{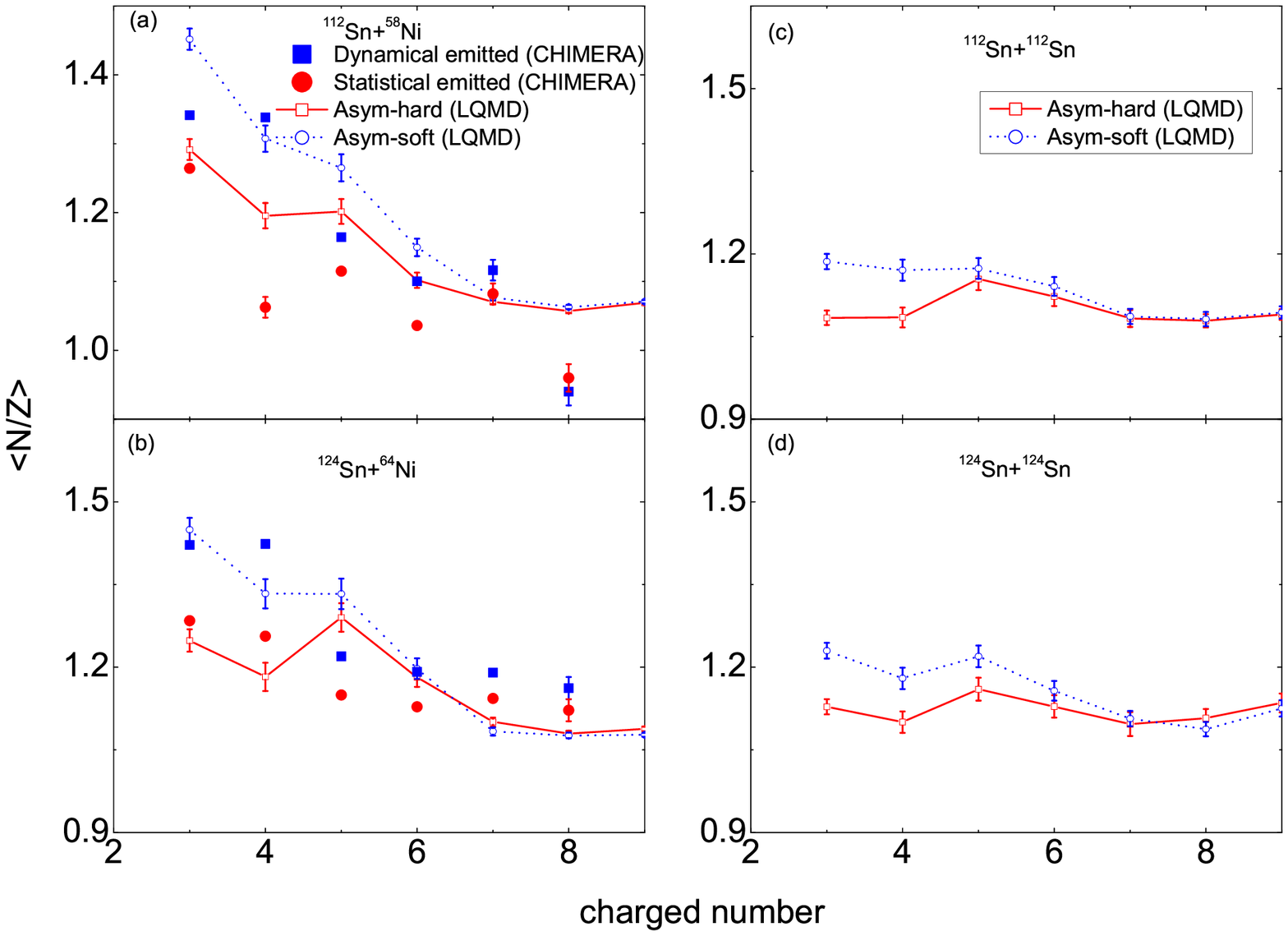}
\caption{(Color online) The average neutron to proton ratio of IMFs emitted within the rapidity range of $|y/y_{proj}|<$0.3 as a function of charged number with the different symmetry energies. The data from CHIMERA detector \cite{Fi12} are shown for comparison.}
\end{figure*}

\begin{figure*}
\includegraphics[width=16 cm]{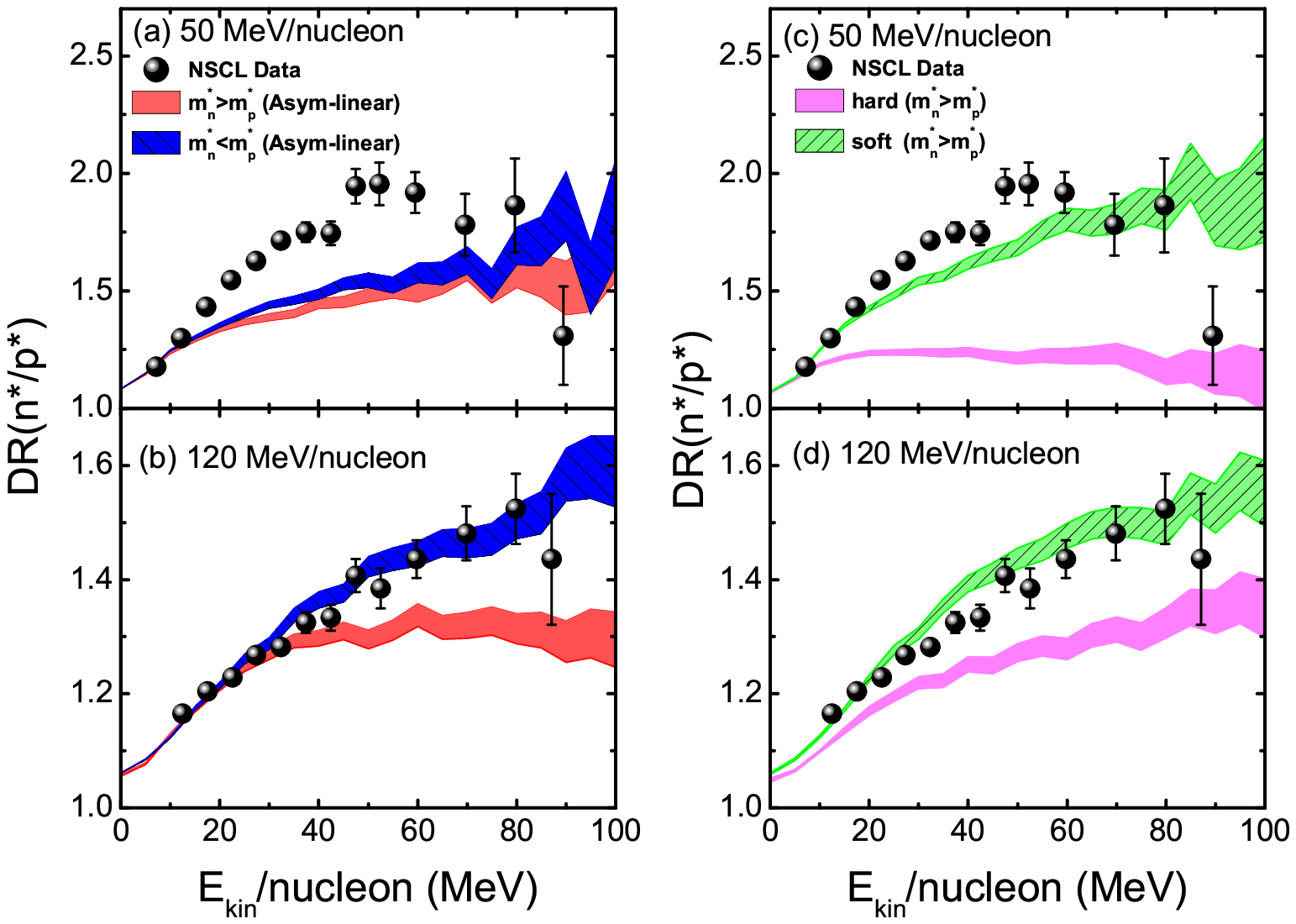}
\caption{(Color online) The double neutron to proton ratios of gas-phase particles in collisions of $^{124}$Sn+$^{124}$Sn over $^{112}$Sn+$^{112}$Sn at the incident energy of 50 MeV/nucleon (upper panels) and 120 MeV/nucleon (lower panels) with different isospin splitting of effective mass and different stiffness of symmetry energy. The available data were measured at NSCL \cite{Co16}.}
\end{figure*}

Besides the fast nucleons being probes of symmetry energy in the dilute matter, the neutron to proton ratio of light intermediate mass fragments (IMFs) could be sensitive to the stiffness of symmetry energy because of the isospin migration from the neck fragmentation \cite{Ba04}. The light IMFs (Z$\leq$10) are measured by the CHIMERA detector at the INFN-LNS Superconducting Cyclotron of Catania (Italy), and emitted preferentially towards the midrapidity domain on a short timescale within 50 fm/c, which can not be entirely described through the decay of the excited projectile-like (PLF) and target-like (TLF) fragments \cite{Fi12}. We constrained the particles emitted from the neck fragmentation within the rapidity range of $|y/y_{proj}|<$0.3. Shown in Fig. 11 is the average neutron to proton ratio of light IMFs in the isotopic reactions of $^{112}$Sn+$^{58}$Ni, $^{124}$Sn+$^{64}$Ni, $^{112}$Sn+$^{112}$Sn and $^{124}$Sn+$^{124}$Sn. The isospin effect is pronounced for the light isotopes, i.e., lithium, beryllium and boron. The soft symmetry energy leads to larger n/p ratio of light IMFs. It is caused from the fact that the heavy fragments tend to the neutron-rich side with the hard symmetry energy. However, the statistical decay leads to the disappearance of the isospin effect in heavy fragments. The light IMFs from the neck fragmentation are emitted preferentially towards the midrapidity domain on a short time scale in comparison to PLFs and TLFs. The isospin ratios depend on the stiffness of the symmetry energy and the effects increase with softening of the symmetry energy, in particular in neutron-rich nuclear reactions. It should be mentioned that the hard symmetry energy leads to the neutron-rich fragment formation in the PLF and TLF regime.

The spectra of the isospin ratios are influenced by both the symmetry energy and the isospin splitting of nucleon effective mass, in particular at the high kinetic energies, which confuse the extraction of the density dependence of symmetry energy. On the other hand, the n/p ratios are also influenced by the Coulomb potential and the detector efficiencies of protons and neutrons in experiments. To eliminate the uncertainties, the double ratios of two isotopic systems would be nice probes for constraining the isospin splitting of nucleon effective mass and the symmetry energy beyond the saturation density from the experimental data. Shown in Fig. 12 is the double ratio spectra in collisions of $^{124}$Sn+$^{124}$Sn over $^{112}$Sn+$^{112}$Sn. The influence of the isospin splitting with a linear symmetry energy (left panels) and the stiffness of symmetry energy with the mass splitting of $m^{*}_{n}>m^{*}_{p}$ (right panels) on the spectra is compared with the new data from the MSU-NSCL (National Superconducting Cyclotron Laboratory) \cite{Co16}. It should be noticed that the effective mass splitting of neutron and proton in nuclear medium is pronounced the kinetic energies above 30 MeV/nucleon and the splitting of $m^{*}_{n}<m^{*}_{p}$ is nicely consistent with the available data at the beam energy of 120 MeV/nucleon. However, the difference of the both mass splittings on the spectra is very small at the beam energy of 50 MeV/nucleon. The symmetry energy effect is obvious and appears at the kinetic energy above 20 MeV/nucleon. A soft symmetry energy ($\gamma_{s}$=0.5) is constrained from the both incident energies \cite{Gu17}. It is concluded that the mass splitting from the isovector momentum dependent interaction potential is negligible in the Fermi-energy heavy-ion collisions. The stiffness of symmetry energy is a dominant quantity on the double ratio spectra. However, the n/p ratios in high-energy heavy-ion collisions are influenced by both the symmetry energy and isospin splitting of effective mass. One should pay attention to some observables independent on the isospin splitting for constraining the density dependence of symmetry energy, such as the charged pion spectra from the high-density domain \cite{Fe12c}. It should be noticed that the fragmentation reactions have been investigated within the antisymmetrized molecular dynamics model for extracting the temperature from the isobaric yield ratios \cite{Ma16}.

\subsection{Pseudoscalar meson production and in-medium effects in heavy-ion collisions}

Hadronic matter is created in high-energy heavy-ion collisions and exists in the compact stars, such as neutron stars. The hadron-hadron interaction in the superdense matter is complicated and varying with the baryon density. The hadron-nucleon potential impacts the ingredients of hadrons in the compact stars. In nuclear reactions, i.e., heavy-ion collisions, antiproton (proton) induced reactions etc, the production and phase-space distribution of particles were modified in nuclear medium. The isospin dependence of the potential and the corrections on threshold energies influence the ratios of isospin particles. Consequently, the extraction of high-density symmetry energy is to be moved from particle production. On the other hand, the yields of particles and bound fragments such as hypernuclides, kaonic nucleus, antiprotonic nucleus is related to the potential of particles in nuclear medium. Dynamics of pseudoscalar mesons ($\pi$, $\eta$, $K$ and $\overline{K}$ ) and hyperons ($\Lambda$ and $\Sigma$) produced in heavy-ion collisions near threshold energies has been investigated within the LQMD transport model. The in-medium modifications on particle production in dense nuclear matter are implemented in the model through corrections on the elementary cross sections and by inclusion of the meson (hyperon)-nucleon potentials, in which the isospin effects are considered. Pion as the lightest meson in the nature can be produced in nucleon-nucleon collisions with a lower threshold energy e.g., E$_{th}$=289 MeV for $\pi^{\pm}$. The stochastic processes in heavy-ion collisions accelerate part of nucleons, which make the particle production at sub-threshold energies. Dynamics of pions produced in heavy-ion collisions has been investigated both in theories and in experiments, which is related to the issues of the high-density symmetry energy and the pion-nucleon interaction in dense matter. The production rate and density profile of $\pi$, $\eta$ and K$^{+}$ produced in central $^{197}$Au+$^{197}$Au collisions at the incident energies of 0.4 and 1.5 \emph{A} GeV are shown in Fig. 13. The reabsorption process retards the pion production towards the subnormal densities. The maximal yields correspond to the densities of 1.0$\rho_{0}$ and 1.6$\rho_{0}$ with and without the pion-nucleon ($\pi$N) scattering, respectively. The average density in the production pion is evaluated by $<\rho>=\int\rho dN/N_{total}$ and $N_{total}$ being the number of total pions, and to be 1.07$\rho_{0}$ and 1.44$\rho_{0}$ for the cases with and without the $\pi$N absorption reactions at the energy of 0.4 \emph{A} GeV, respectively. Therefore, pions produced in heavy-ion collisions provide the information of phase diagram around saturation densities. The production of $\eta$ is similar to the pion dynamics. However, K$^{+}$ is created within a narrower time range at suprasaturation densities. The average density of the total kaons is around 2$\rho_{0}$ at the energy of 1.5 \emph{A} GeV. One might evaluate the $\pi^{-}/\pi^{+}$ ratio squeezed out the reaction zone for extracting the high-density symmetry energy.

\begin{figure*}
\includegraphics[width=16 cm]{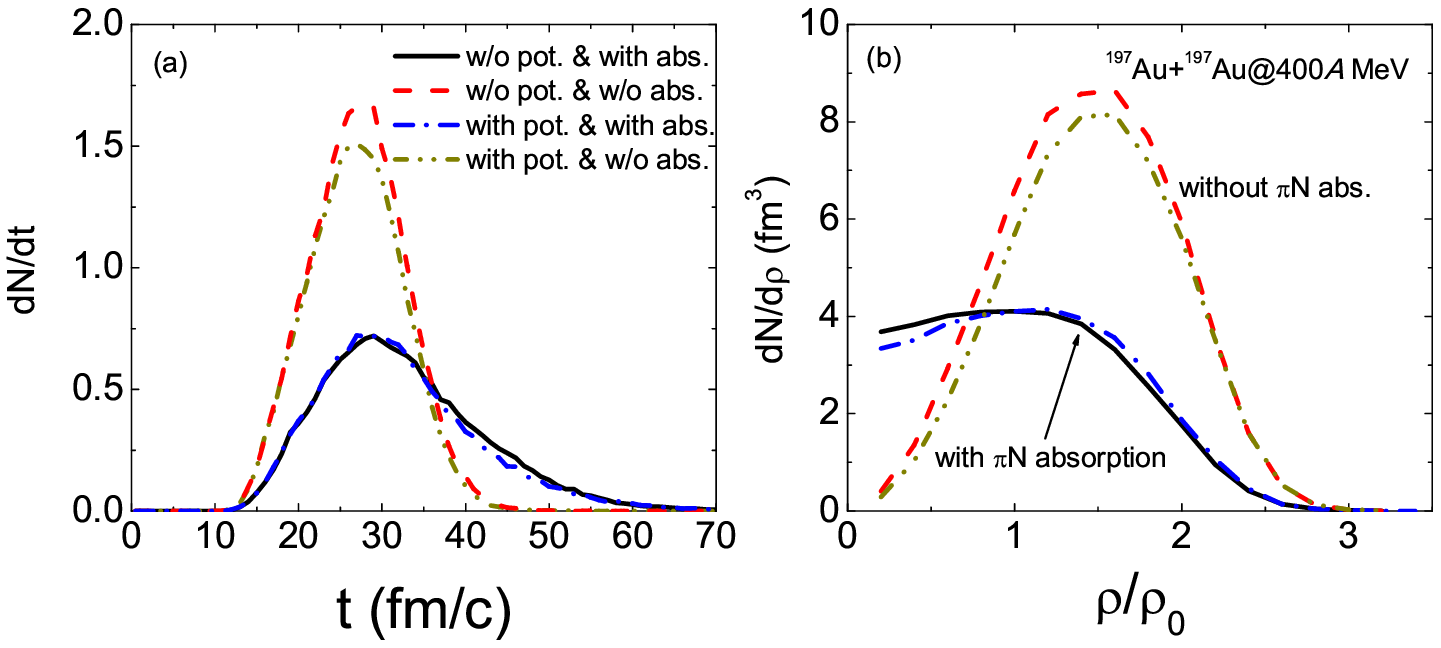}
\includegraphics[width=16 cm]{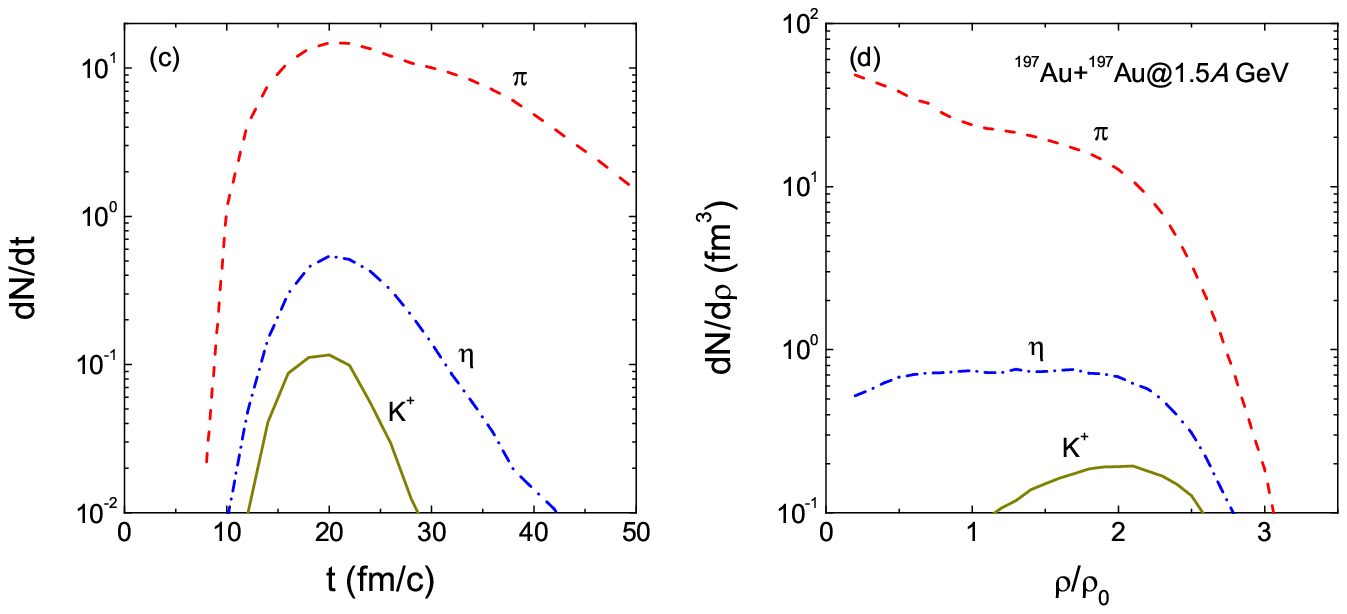}
\caption{(Color online) Evolution of production rate and density profiles of particles in collisions of $^{197}$Au+$^{197}$Au.}
\end{figure*}

Dynamics of pseudoscalar mesons in heavy-ion collisions near threshold energies provides the isospin effect of nucleon force and the in-medium properties of u, d and s quark in dense matter. The strange particles are produced at the supra-saturation densities and emitted at the early stage of the phase diagram comparing with the pion production. Strangeness exchange process retards the saturation of $\Lambda$ and $\Sigma$ production \cite{Fe10c}. Higher incident energy and central collisions enhance the domain of the high-density phase diagram, hence lead to the increase of the strangeness multiplicity. A larger high-density region of compressed nuclear matter is formed in heavy collision systems comparing with the light systems, which increases the strangeness production. The total multiplicity of particles produced in central $^{40}$Ca+$^{40}$Ca collisions are calculated shown in Fig. 14. The $\eta$ and strange particles (K, $\overline{K}$, $\Lambda$, $\Sigma$) are enhanced owing to the $\pi$ induced reactions, such as $\pi N\rightarrow N\ast(1535)$, $\pi N\rightarrow KY$ and $\pi N\rightarrow NK\overline{K}$. The available data from FOPI collaboration for pions \cite{Re07} and from TAPS collaboration for etas \cite{Av03} can be well reproduced. It should be mentioned that the inclusion of the KN potential in the model leads to about 30$\%$ reduction of the total kaon yields in the subthreshold domain \cite{Fe13a}. The competition of the $\pi (\eta)$N and KN potentials results in a bit of increase of kaon production because of the contribution of $\pi (\eta)N\rightarrow KY$.

\begin{figure*}
\includegraphics[width=16 cm]{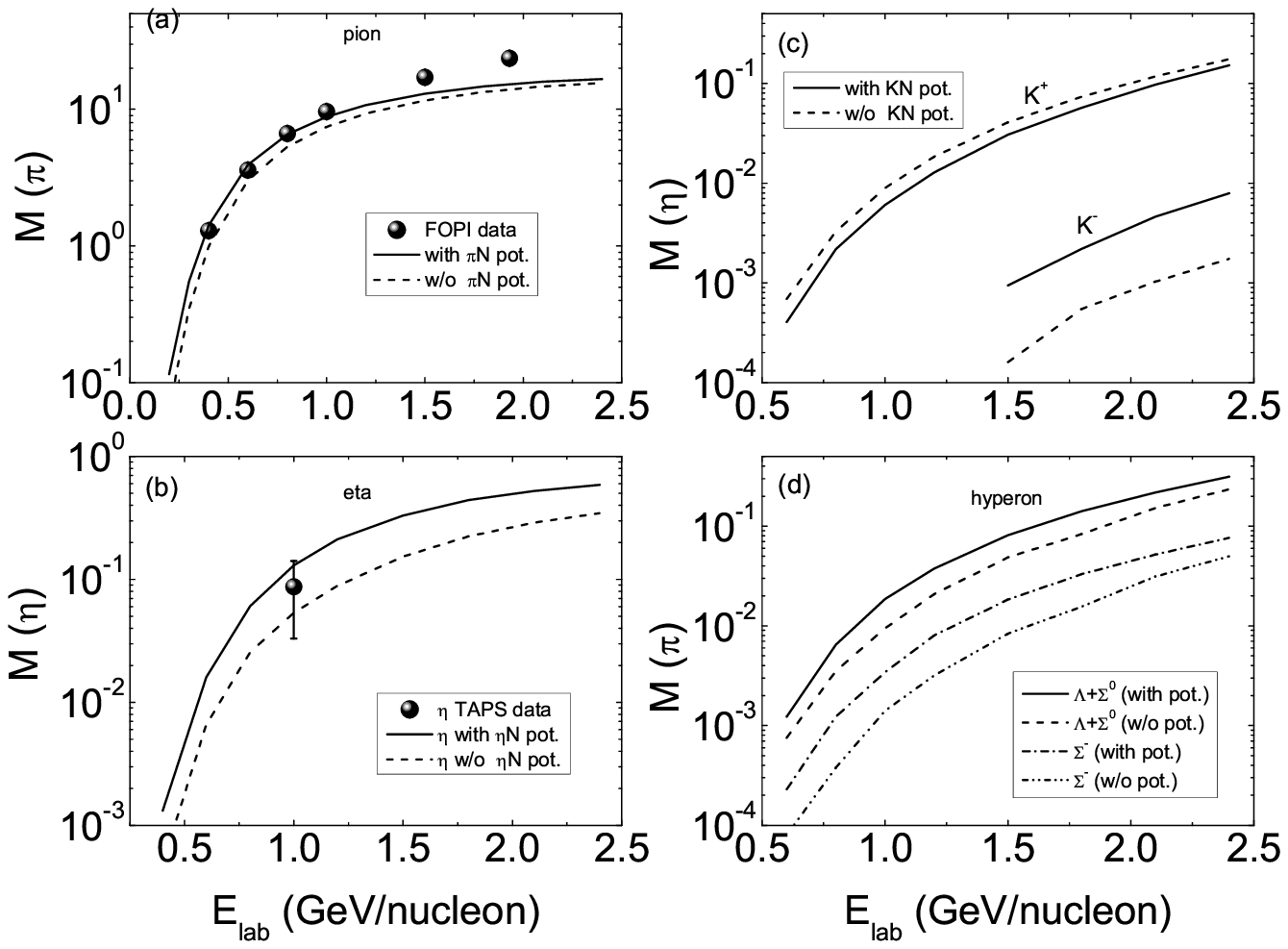}
\caption{\label{fig:wide} Excitation functions of particles produced in central $^{40}$Ca+$^{40}$Ca collisions. The available data from FOPI collaboration for pions \cite{Re07} and from TAPS collaboration for etas \cite{Av03} are shown for comparison.}
\end{figure*}

\begin{figure*}
\includegraphics[width=16 cm]{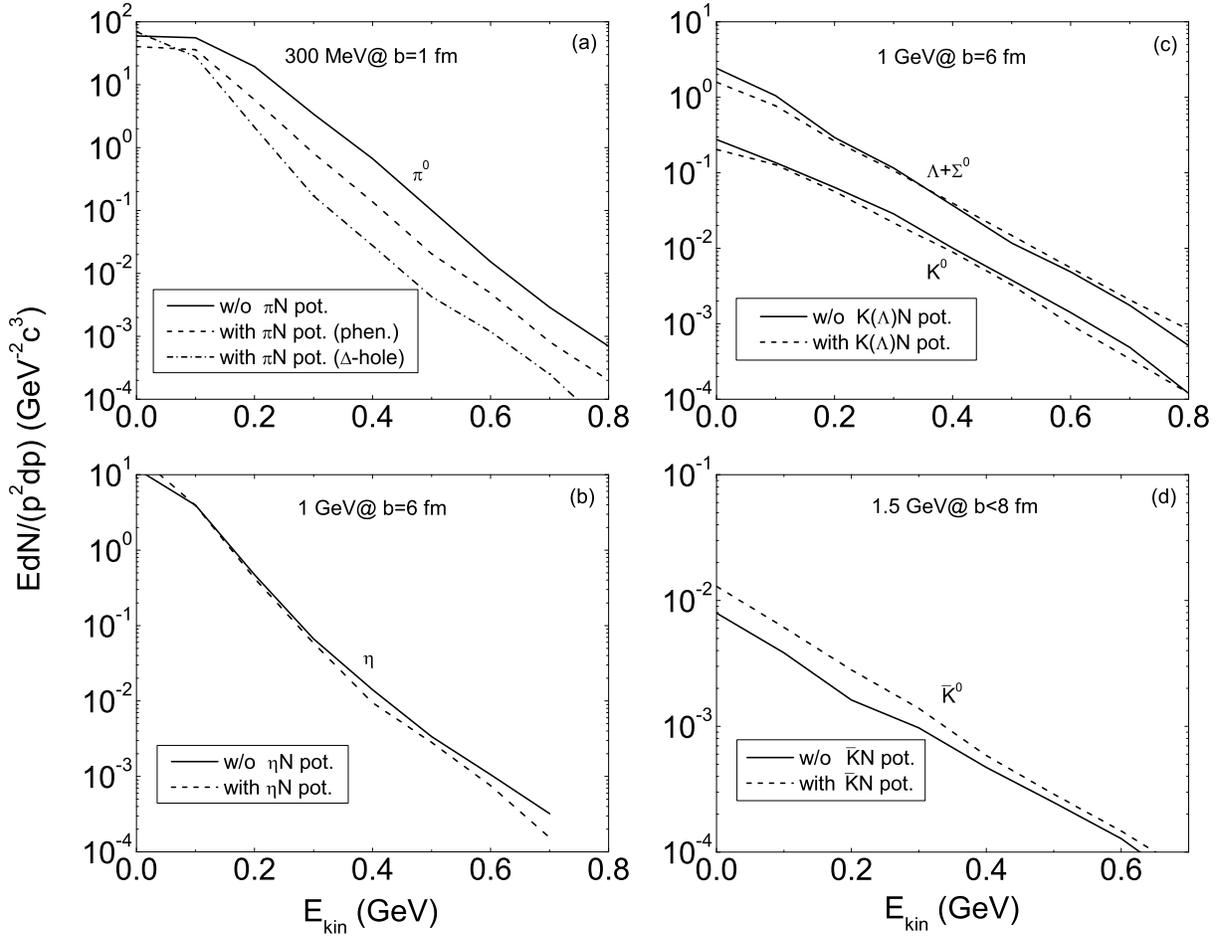}
\caption{Inclusive spectra of the neutral particles produced in $^{197}$Au+$^{197}$Au collisions and in-medium effect on the particle production \cite{Fe15a}.}
\end{figure*}

\begin{figure*}
\includegraphics[width=16 cm]{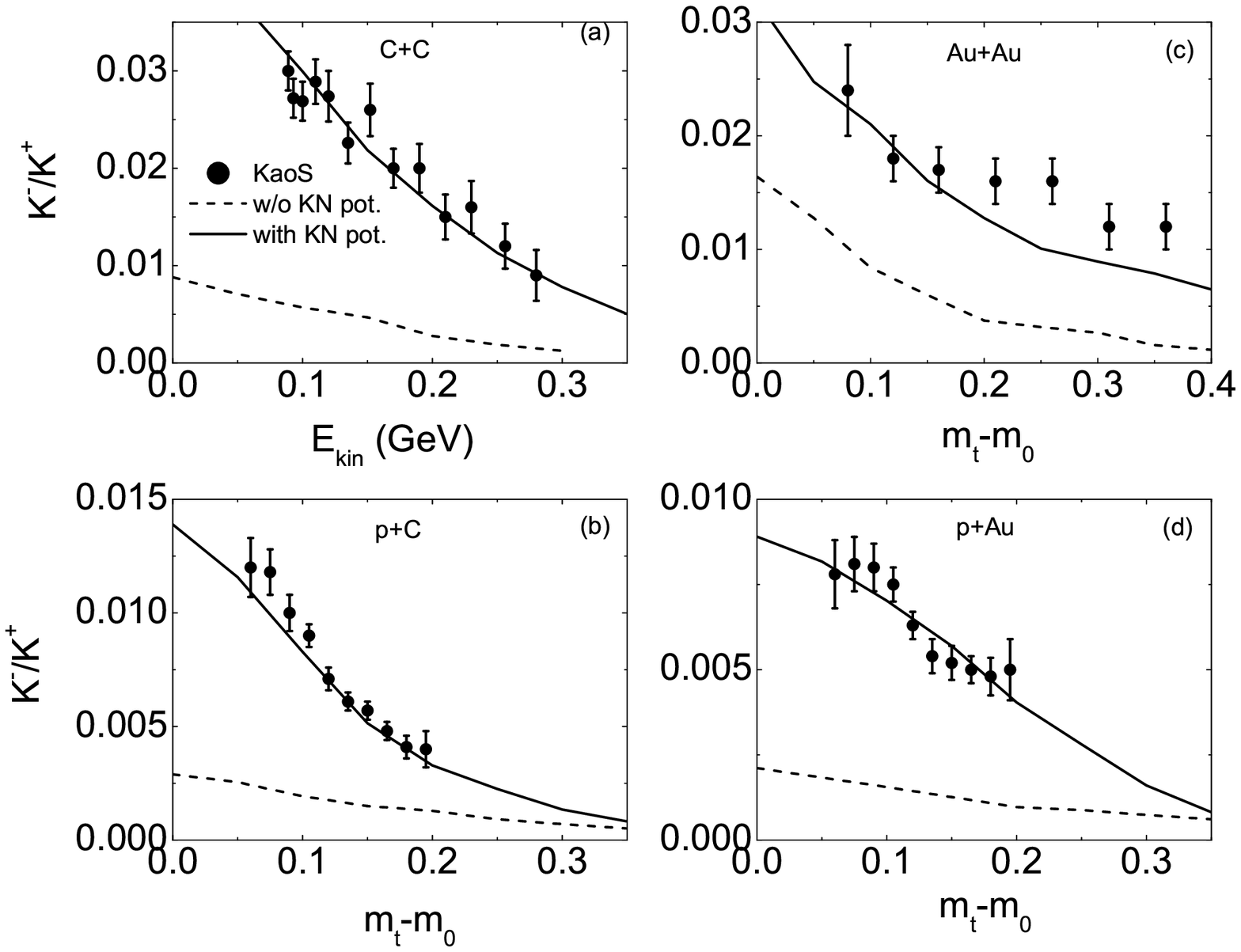}
\caption{The ratio of K$^{-}$/K$^{+}$ as a function of transverse mass (kinetic energy) in collisions of $^{12}$C+$^{12}$C and protons on $^{12}$C and $^{197}$Au at the beam energies of 1.8\emph{A} GeV and 2.5 GeV, respectively. The reaction of $^{197}$Au+$^{197}$Au at the energy of 1.5\emph{A} GeV \cite{Fe14b}. The experimental data are taken from the KaoS collaboration \cite{La99,Fo03,Sc06}}
\end{figure*}

To investigate the high-density symmetry, the isospin observables emitted from the high-density reaction zone are expected. However, the dynamics of particles produced in the high-density reaction domain is distorted in nuclear medium. The in-medium corrections on particle production in heavy-ion collisions are the elementary cross section modifications and the dynamical evolutions by the mean-field potentials \cite{Fe15a}. The interaction potential of the strange particle and nucleon is of significance in the formation of a hypernucleus, the core structure of a neutron star, etc. However, it is not well understood up to now, in particular, in the dense nuclear matter. Consequently, the KN potential reduces the kaon production at midrapidities and at high transverse momenta. However, an opposite contribution of the $\eta$N potential is obtained because of the attractive interaction of eta and nucleon in nuclear medium. The hyperon-nucleon interaction is negligible for $\Lambda$ dynamics. Shown in Fig. 15 is the inclusive spectra of $\pi^{0}$, $\eta$, $K^{0}$, $\overline{K}^{0}$ and neutral hyperons ($\Lambda$+$\Sigma^{0}$) in $^{197}$Au+$^{197}$Au collisions. Similar to the transverse momentum spectra, the eta-nucleon, kaon-nucleon and hyperon-nucleon potentials weakly impact the particle emission. However, the interaction of pions (antikaons) and nucleons contributes the pion (antikaon) dynamics, i.e., increasing the antikaon production at low kinetic energies. More pronounced information of the in-medium effects has been investigated from the K$^{-}$/K$^{+}$ spectrum in heavy-ion collisions. A deeply attractive $K^{-}N$ potential being the value of -110$\pm$15 MeV was obtained at saturation density and weakly repulsive $K^{+}N$ potential has been concluded to be 25$\pm$10 MeV from heavy-ion collisions \cite{Li97,Ca99,Fu06,Ha12}. The in-medium modifications influence the kaon production and dynamical emissions in phase space, i.e., inclusive spectrum, collective flows etc. We compared the effects of kaon(antikaon)-nucleon potentials in the $^{12}$C+$^{12}$C reaction as a function of kinetic energy and in collisions of $^{197}$Au+$^{197}$Au and proton on $^{12}$C and $^{197}$Au versus the transverse mass ($m_{t}=\sqrt{p_{t}^{2}+m_{0}^{2}}$ with $p_{t}$ being the transverse momentum and the mass of kaon (antikaon) $m_{0}$) as shown in Fig. 16 \cite{Fe14b}. The experimental data from KaoS collaboration \cite{La99,Fo03,Sc06} can be nicely reproduced within inclusion of the mean-field potentials for kaons and antikaons, in which a weakly repulsive $KN$ potential of the order of 28 MeV and a deeply attractive $\overline{K}N$ potential of -100 MeV at normal nuclear density are used in the LQMD model. The attractive $\overline{K}N$ potential reduces the threshold energies associated with enhancing $K^{-}$ production. However, the $KN$ potential leads to an opposite contribution for the $K^{+}$ emission. The value of $K^{-}/K^{+}$ ratio is more sensitive to the optical potentials because of the opposite interactions of kaon and antikaon in nuclear medium. The $K^{-}$-bound state and kaon condensation in dense baryonic matter are being searched in experiments.

\begin{figure*}
\includegraphics[width=16 cm]{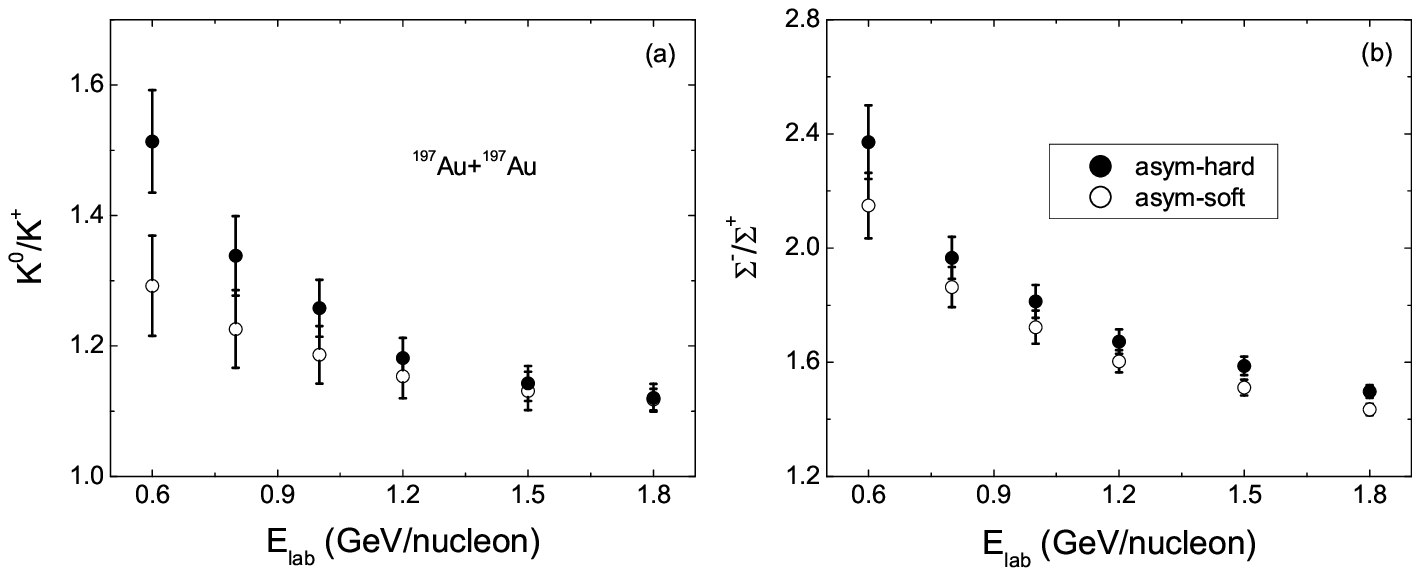}
\caption{Excitation functions of the isospin ratios of K$^{0}$/K$^{+}$ and $\Sigma^{-}/\Sigma^{+}$ in the $^{197}$Au + $^{197}$Au reaction for the hard and soft symmetry energies, respectively.}
\end{figure*}

The high-density matter larger than normal nuclear density can be formed in high-energy heavy-ion collisions, where strange particles are produced. The influence of the in-medium potential by distinguishing isospin effect and the stiffness of symmetry energy on strangeness production can be observed from the emission of isospin pairs. Shown in Fig. 17 is a comparison of the energy dependence of the yield ratios of K$^{0}$/K$^{+}$ and $\Sigma^{-}/\Sigma^{+}$ produced in central $^{197}$Au+$^{197}$Au collisions. It should be noticed that the KN potential by distinguishing isospin effect increases the K$^{0}$/K$^{+}$ ratio and also enlarges the influence of symmetry energy on isospin ratio in comparison to the in-vacuum and the nonisospin cases, in particular in the domain of deep subthreshold energies. The effect of symmetry energy on the isospin ratios disappears at high incident energy. The inclusion of the KN potential leads to a reduction of the total kaon yields. The contribution of LF is very small on the kaon production, but changes the kaon distribution in momentum space \cite{Fe13a,Fe13b}. Calculations of the double ratio excitations of K$^{0}/$K$^{+}$ taken from two systems of $^{96}$Ru+$^{96}$Ru and $^{96}$Zr+$^{96}$Zr presented that the symmetry energy effect is very weak over the whole energy range \cite{Fe13b}. In analogue to the isospin ratio of kaons, the effect of symmetry energy on the $\Sigma^{-}/\Sigma^{+}$ excitation function is pronounced in the region of the deep threshold energies. Overall, a hard symmetry energy leads to a lower isospin ratio in neutron-rich nuclear collisions. The $\Sigma^{-}/\Sigma^{+}$ ratio is larger than the value of $K^{0}/K^{+}$ in the whole energy regime. Production of $\Sigma$ increases dramatically with the incident energy and the isospin effects of the yields for charged $\Sigma$ are obvious with decreasing the energy. In despite of part of sigma can be absorbed in nuclear matter through the reaction of $\pi \Sigma \rightarrow \overline{K}B$. Most of $\Sigma$ are produced in the high-density region in heavy-ion collisions, which can be used to probe the information of the high-density phase diagram of nuclear matter. The KN and $\Sigma$N potentials are of importance on the kaon and $\Sigma$ emissions in phase space. The isospin ratio of $K^{0}/K^{+}$ depends on the kaon potential at high transverse momenta and a flat spectrum appears after inclusion of the KN potential. The $K^{0}/K^{+}$ and $\Sigma^{-}/\Sigma^{+}$ ratios of neutron-rich heavy systems in the domain of subthreshold energies are sensitive to the stiffness of nuclear symmetry energy, which are created in the high-density domain in heavy-ion collisions and might be the promising probes to extract the high-density information of symmetry energy.

\subsection{Hypernucleus formation in heavy-ion collisions}

\begin{figure*}
\includegraphics[width=16 cm]{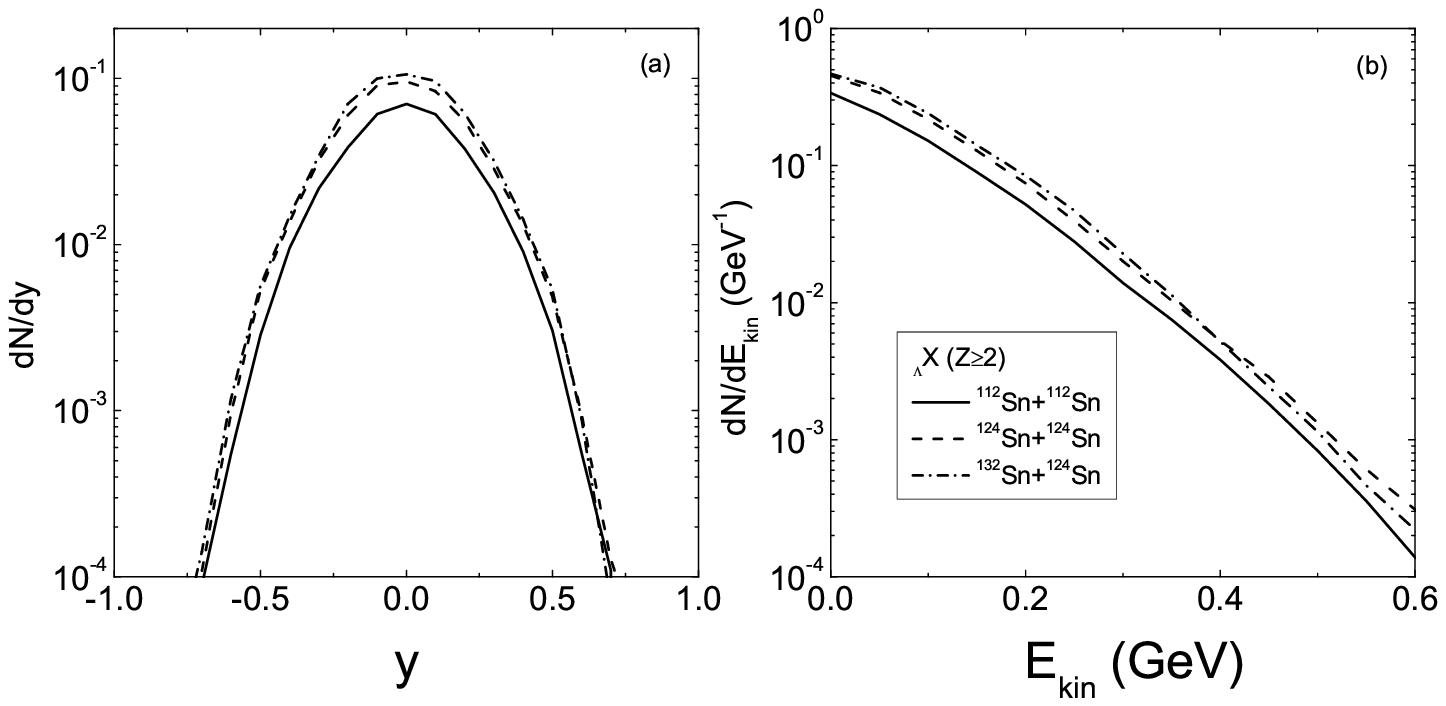}
\caption{Rapidity and kinetic energy distributions of hypernuclides with the charged numbers of Z$\geq$2 in the isotopic reactions.}
\end{figure*}

\begin{figure*}
\includegraphics[width=16 cm]{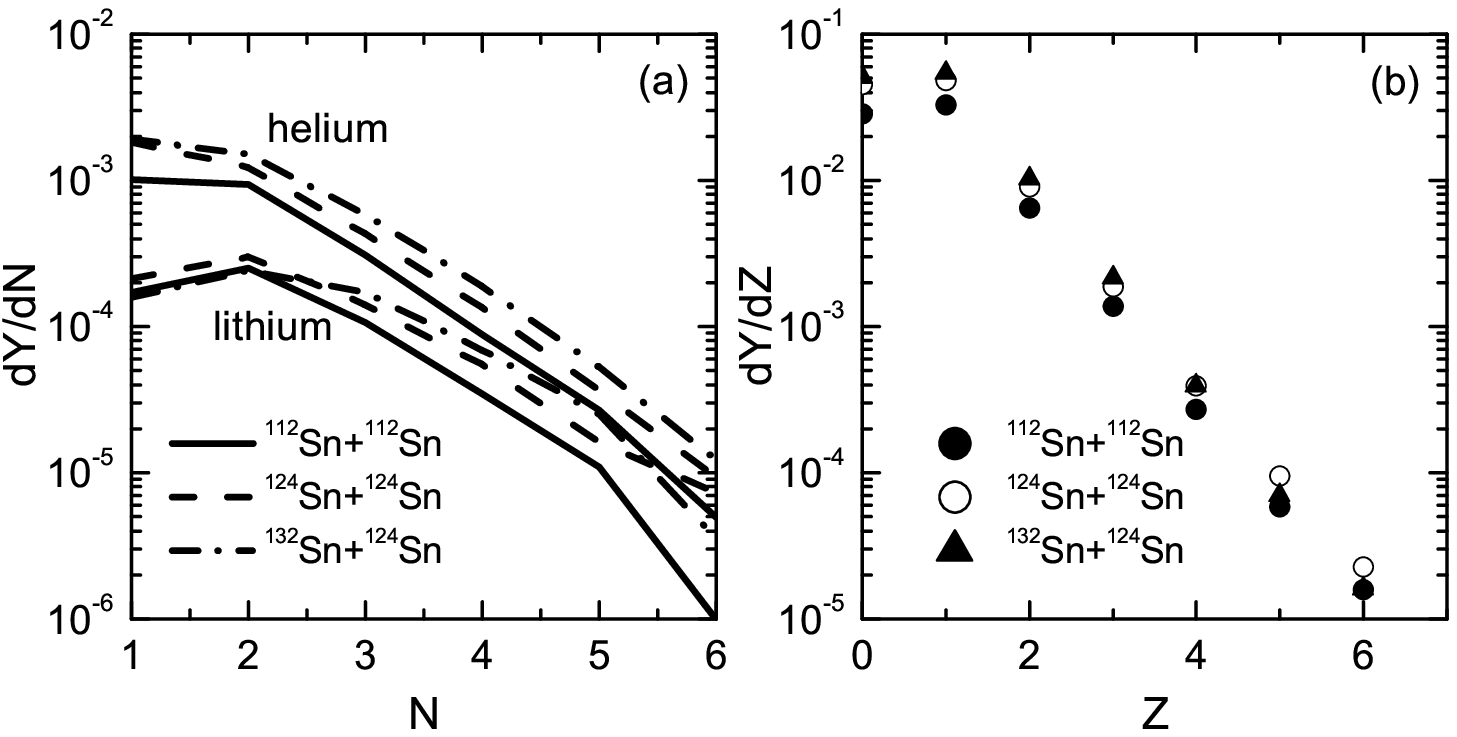}
\caption{Isotopic and charge distributions of $\Lambda$-hypernuclides produced in the $^{132,124,112}$Sn + $^{112}$Sn/$^{124}$Sn collisions at the incident energy of 2$\emph{A}$ GeV.}
\end{figure*}

Formation mechanism of fragments with strangeness in high-energy heavy-ion collisions has been investigated within the LQMD transport approach combined with the statistical model (GEMINI). The production and dynamics of hyperons is described within the LQMD model. A coalescence model is used for constructing the primary fragments and the hyperon capture by residual nucleons. The combined approach is used to describe the formation of hyperfragments. The production of hypernuclei is associated to the hyperon production, hyperon-nucleon and hyperon-hyperon interactions, capture of hyperons by nucleonic fragments, and decay of excited hyper-fragments. The investigation of hypernucleus properties is an essential way for extracting the in-medium information of hyperons. To form the hyperfragments in heavy-ion collisions, the incident energy is chosen to be enough for creating hyperons, but not too high so that the hyperons being captured by surrounding nucleons. At the near threshold energies, the reaction channels of $BB \rightarrow BYK$ and $\pi(\eta)B \rightarrow KY$ dominate the hyperon production. Usually, the hyperons are created in the domain of the dense nuclear medium. The hyperon-nucleon potential impacts the hyperon dynamics and hyperfragment formation. Calculations from a statistical model gave that the beam energy regime of 3-5 GeV/nucleon is available for producing hypernuclei \cite{An11}. The production and structure studies of neutron-rich and even double-strangeness hypernuclides have been planned at the high intensity heavy-ion facility (HIAF) in the future. Shown in Fig. 18 is the rapidity and kinetic energy spectra of $\Lambda$-hypernuclides with the charged numbers of Z$\geq$2 in collisions of $^{112}$Sn+$^{112}$Sn, $^{124}$Sn +$^{124}$Sn and $^{132}$Sn +$^{124}$Sn at the incident energy of 2 GeV/nucleon and within the collision centrality bin of 0-8 fm in the center of mass frame. The hyperfragments are formed within a narrower rapidity domain and at the less kinetic energies in comparison to hyperons in Fig. 15. The phase-space distributions are available for arranging the detector setup in experiments. The isotopic and charge distributions of $\Lambda$-hyperfragments are also calculated as shown in Fig. 19. It is obvious that the neutron-rich systems are favorable for producing the more neutron-rich hyperfragments and the production rate decreases drastically with the charged number. The medium and heavy hyperfragments might be created with the peripheral collisions, but having very small cross sections owing to the less production number of hyperons. The strangeness physics opens a new window in nuclear physics, besides the hypernuclide properties, and also being the main ingredients in the dense baryonic matter, such as neutron star etc. Accurate description of the hypernucleus formation in heavy-ion collisions is still in progress, i.e., the quantal clustering hyperfragments, the multi-mode decays of excited hyperfragments, the formation of exotic hypernucleus ($^{3}_{\Lambda}n$, $^{5}_{\Lambda}n$) etc.

\section{Nuclear dynamics induced by hadrons}

The nuclear matter formed in heavy-ion collisions is compressed and the nuclear density varies with the time-space evolutions and the incident energy. The spinodal and chemical instabilities complicate the nuclear fragmentation and particle emission in heavy-ion collisions. However, the nuclear reactions induced by hadrons have advantages on many topical issues, i.e., the in-medium properties of hadrons around the saturation density, the highly excited nucleus, the production of hypernucleus etc.

\subsection{Particle production in proton-nucleus collisions}

The nuclear dynamics induced by protons has many interest aspects, in particular, the spallation reactions on heavy targets, the in-medium properties of hadrons, the hypernucleus formation etc. Microscopic description on the issues is quite necessary. It has been found the strangeness production is strongly suppressed in proton-induced reactions in comparison to heavy-ion collisions \cite{Fe14b}. The kaon-nucleon and antikaon-nucleon potentials change the structures of rapidity and transverse momentum distributions and the inclusive spectra in proton-nucleus collisions. Shown in Fig. 20 is the temporal evolution of pions, resonances, $\eta$ and strange particles produced in collisions of proton on $^{40}$Ca at the momentum of 3 GeV/c. The pions are dominant products and the secondary collisions of
$\pi(\eta)N \rightarrow KY$ is negligible for the hyperon production. Once the particles are created, they go away the surrounding nucleons rapidly and the shadowing effect is very weak in the proton induced reactions. The strange particles are mainly from the direct process of the channel $NN(\Delta) \rightarrow NYK$. The phase-space distributions of particles produced in proton-nucleus collisions are distorted by the surrounding nucleons although the secondary collisions are slight. Therefore, it provides a good opportunity for studying the optical potentials around the normal nuclear density and the exotic bound states with the proton induced reactions, i.e., $\pi$N, $\eta$N, YN etc.

\begin{figure*}
\includegraphics[width=16 cm]{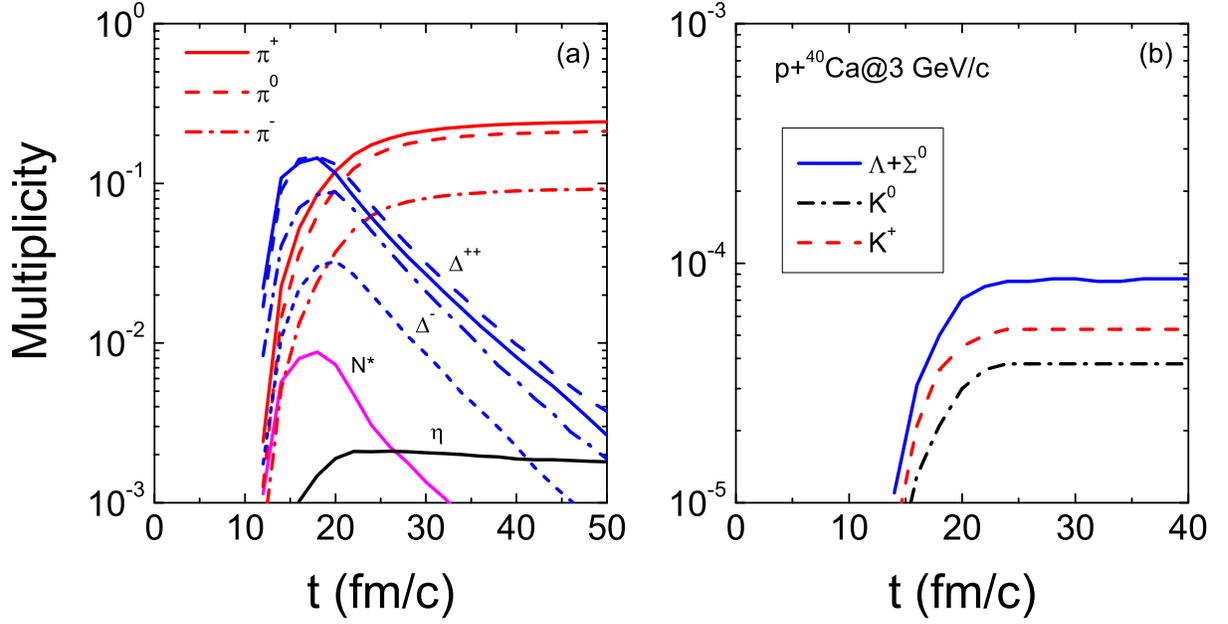}
\caption{Temporal evolution of particle production in collisions of proton on $^{40}$Ca at the momentum of 3 GeV/c.}
\end{figure*}

\begin{figure*}
\includegraphics[width=16 cm]{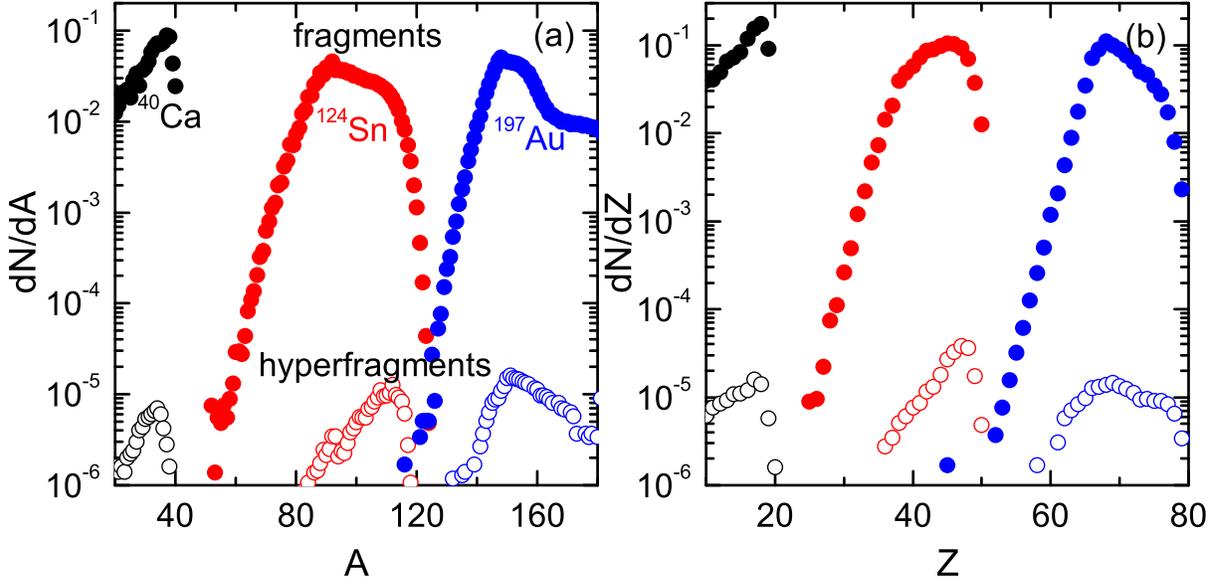}
\caption{Production of nucleonic fragments and hyperfragments in collisions of proton on $^{40}$Ca, $^{124}$Sn and $^{197}$Au at the momentum of 5 GeV/c, respectively.}
\end{figure*}

It has been found that the kaon- (antikaon-) nucleon potential plays a significant role on the strangeness production
and dynamical emission in phase space, which reduces (enhances) the kaon (antikaon) yields and is pronounced when the incident energy is close to the threshold values. Although the very low hyperon yields produced in proton-nucleus collisions, the hyperfragments could be formed because the hyperons are created inside the target nucleus. The fragments and hyperfragments are calculated as shown in Fig. 21 with the 5 GeV/c protons on the targets of $^{40}$Ca, $^{124}$Sn and $^{197}$Au, respectively. Roughly, the $\Lambda$-hyperfragments are produced with the yields of the 4-order magnitude lower than the nucleonic fragments. The fragments tend to be formed in the target-like region. It provides the possibility for studying the delayed fission of heavy hypernuclide with the high-energy protons.

\subsection{Meson induced nuclear reactions}

The nuclear dynamics induced by charged mesons has been attracted much attention on the aspects of the charge-exchange reactions, nuclear fragmentation, hypernuclide production etc. On the other hand, the meson-nucleus collisions have advantages to investigate the energy dissipation mechanism, the meson-nucleon interaction, resonance properties in nuclear medium around the saturation density. Shown in Fig. 22 is the comparison of the nucleonic fragments, hyperfragments and free $\Lambda$  in the $K^{-}$+$^{40}$Ca reaction at an incident momentum of 2 GeV/c. A similar structure of hyperfragments and free $\Lambda$ is found. The strangeness exchange reaction of $K^{-}N \rightarrow \pi  Y$ dominates the hyperon production. Once the $\Lambda$ is created in the nucleus, a large probability is captured to form a hypernucleus. The free $\Lambda$ hyperons and hyperfragments have the narrower rapidity regime and the lower kinetic energies. The nuclear fragmentation is still pronounced owing to the decay of highly excited target nucleus.

\begin{figure*}
\includegraphics[width=16 cm]{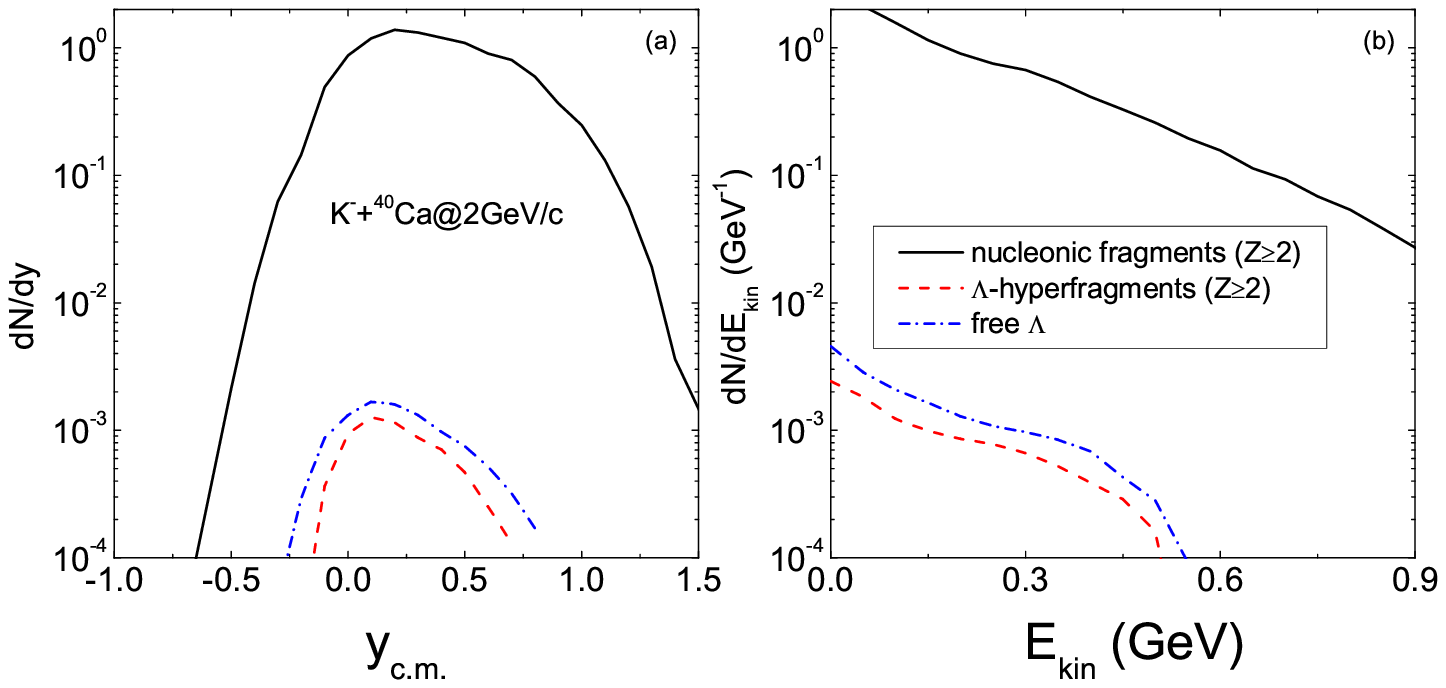}
\caption{Rapidity and kinetic energy distributions of nucleonic fragments, hyperfragments and free $\Lambda$ in the $K^{-}$+$^{40}$Ca reaction at an incident momentum of 2 GeV/c.}
\end{figure*}

\begin{figure*}
\includegraphics[width=16 cm]{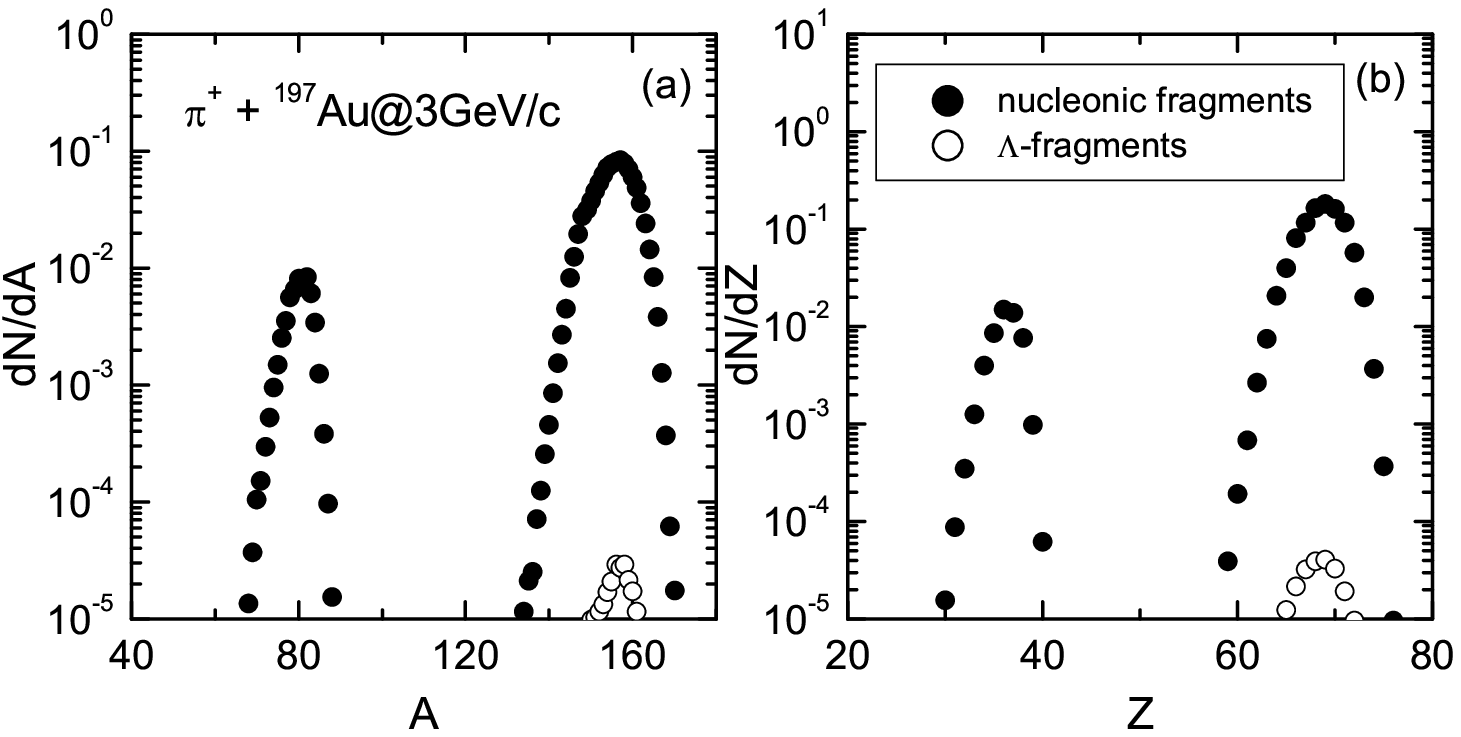}
\caption{Nucleonic fragment and hyperfragment production in collisions of $\pi^{+}$ on $^{197}$Au at the momentum of 3 GeV/c.}
\end{figure*}

Nuclear reactions induced by pions provide the opportunity to study the pion-nucleon interaction, the charge-exchange reactions, decay modes of highly excited nucleus, hypernucleus formation and the in-medium properties of $\Delta$ resonance. Recently, the bound state of the neutron-rich $\Lambda$-hypernucleus $_{\Lambda}^{6}$H was studied with double charge-exchange reaction $^{6}$Li($\pi^{-}$,K$^{+}$)X at a beam momentum of 1.2 GeV/c at J-PARC \cite{Ho17}. The nuclear fragmentations and the charge exchange reactions in pion-nucleus collisions near the $\Delta$(1232) resonance energies has been investigated within the LQMD model \cite{Fe16a}. It is found that the fragments tend to be formed along the $\beta$ stability line and the fission reactions take place for heavy targets. The relative motion energy is deposited in the nucleus via the pion-nucleon collisions for the pion induced reactions. The dissipated energy weakly depends on the incident pion energy, which is available the hypernuclide formation. Shown in Fig. 23 is the distributions of fragment and hyperfragment produced in collisions of $\pi^{+}$ on $^{197}$Au at the momentum of 3 GeV/c. The medium fragments are from the fission of target-like nuclei. It is noticed that the $\Lambda$- hyperfragments are formed in the target-like region.

\begin{figure}
\includegraphics[width=8 cm]{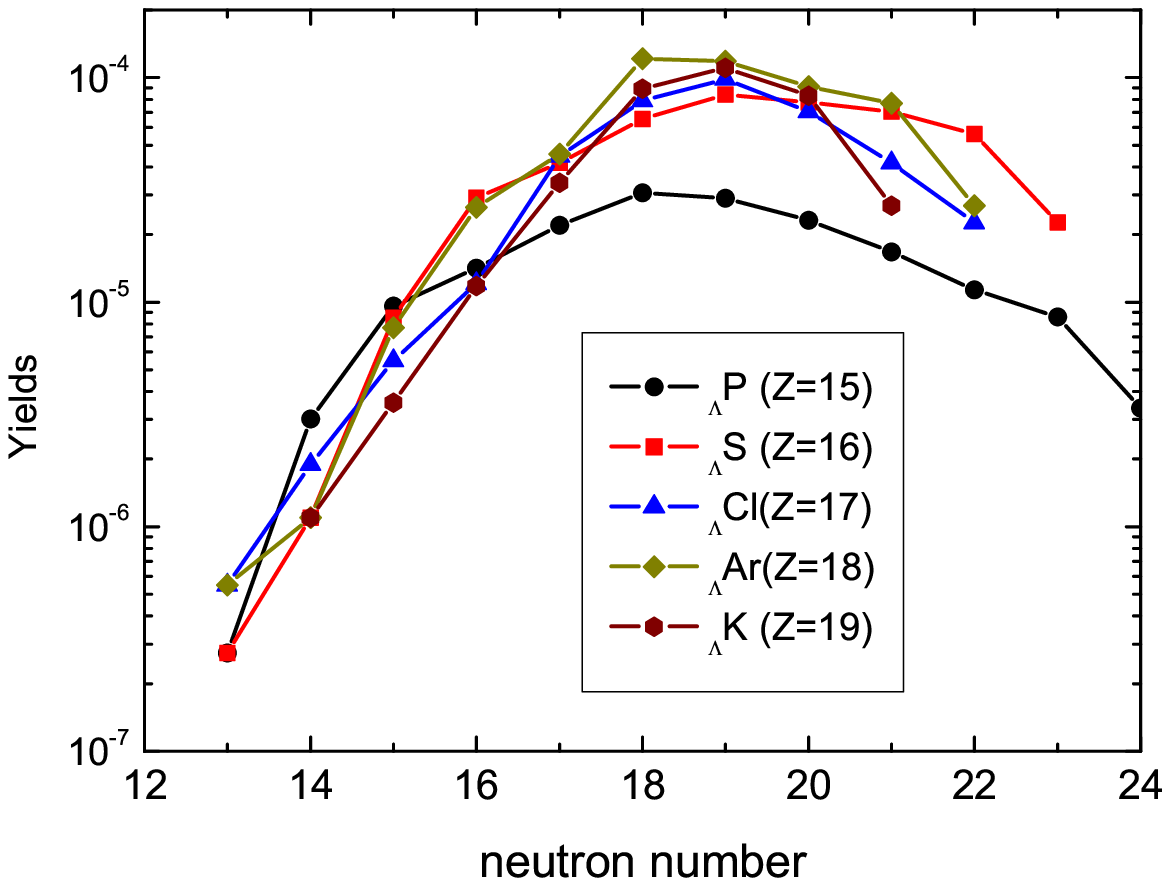}
\caption{Isotopic distributions of hyperfragments in collisions of $K^{-}$ on $^{40}$Ca at the momentum of 2 GeV/c.}
\end{figure}

\begin{figure*}
\includegraphics[width=16 cm]{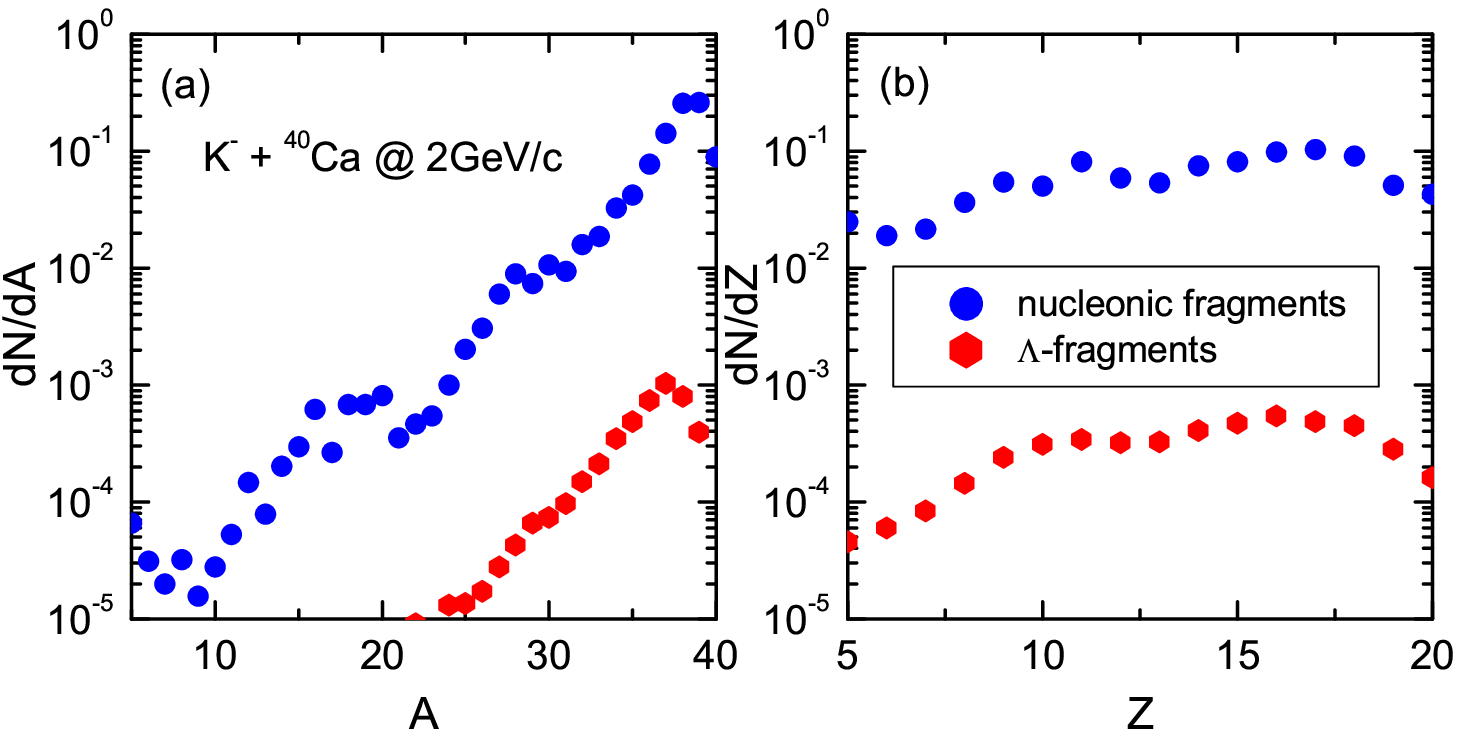}
\caption{Comparison of nucleonic fragments and hyperfragments in collisions of $K^{-}$ on $^{40}$Ca at the momentum of 2 GeV/c.}
\end{figure*}

The hyperon does not obey the Pauli principle and can occupy the same quantum state with nucleons. On the other hand, the attractive hyperon-nucleon interaction enable for binding the very neutron-rich hypernucleus and even the two-neutron-$\Lambda$ or four-neutron-$\Lambda$ state. The isotopic distributions of hyperfragments in collisions of $K^{-}$ on $^{40}$Ca at the momentum of 2 GeV/c are shown in Fig. 24. The maximal yields are located around the neutron number of N=19. A comparison of nucleonic fragments and hyperfragments is shown in Fig. 25. The decay of primary hyperfragments enables the nucleon emissions and lightens the hypernuclei. The hyperfragment yields are reduced the two order magnitude than the nucleonic fragments, which are available for hypernucleus formation with the strangeness exchange reactions in comparison to heavy-ion collisions and $\pi$ induced reactions.

\subsection{Hypernucleus formation in antiproton-nucleus collisions}

Since the first evidence of antiprotons was found in 1955 at Berkeley in collisions of protons on copper at the energy of 6.2 GeV by Chamberlain, Segr\`{e}, Wiegand and Ypsilantis \cite{Ch55}, the secondary beams of antiprotons were produced at many laboratories, such as CERN, BNL, KEK etc \cite{Ch57,Ag60,Le80,Ea99}. The stochastic cooling method provides the possibility for storing the antiprotons produced in proton-nucleus collisions. The particles $W^{\pm}$ and $Z^{0}$ were found for the first time with the high energy protons colliding the stored antiprotons at CERN \cite{Ua83a,Ua83b}. On the other hand, the antiproton-nucleus collisions are motivated to many interesting issues, i.e., charmonium physics, strangeness physics, antiprotonic atom, symmetry, in-medium properties of hadrons, cold quark-gluon plasma, highly excited nucleus etc \cite{Ra80,Am02}. Recently, the antiproton-antiproton interaction was investigated by the STAR collaboration in relativistic heavy-ion collisions \cite{St15}. The low-energy antiprotons usually annihilate at the nucleus surface because of the large absorption cross section. The huge annihilation energy are available for producing the 2-6 pions. The subsequent processes are complicated and also associated with the multiple pion-nucleon interaction, which result in the fragmentation of target nucleus and the preequilibrium emissions of complex particles. The localized energy is deposited in the nucleus with an excitation energy of several hundreds of MeV. The hot nucleus proceeds to the explosive decay via multifragmentation process or the sequential particle evaporation. On the other hand, the collisions of the antiproton and secondary particles with surrounding nucleons lead to the pre-equilibrium particle emissions, which are related to the scattering cross sections of each reaction channels, antiproton-nucleon interaction, particle-nucleon potentials, density profile of target nucleus. The unexpected large neutron yields produced by stopped antiprotons in nuclei were reported in the LEAR experiments \cite{Pol95}.

\begin{figure*}
\includegraphics[width=16 cm]{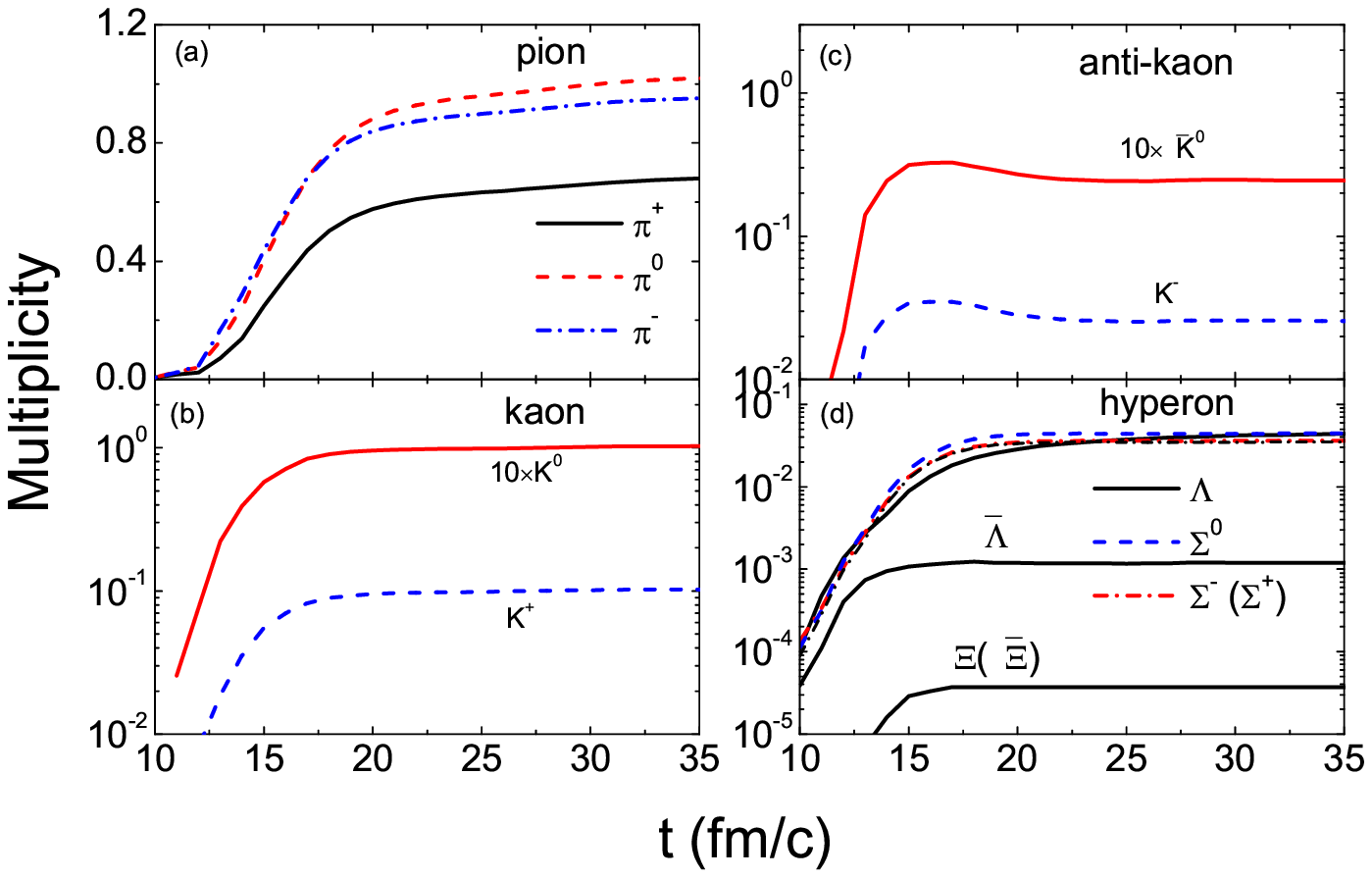}
\caption{(Color online) Temporal evolution of particle production in collisions of $\overline{p}$ on $^{40}$Ca at incident momentum of 4 GeV/c.}
\end{figure*}

\begin{figure*}
\includegraphics[width=16 cm]{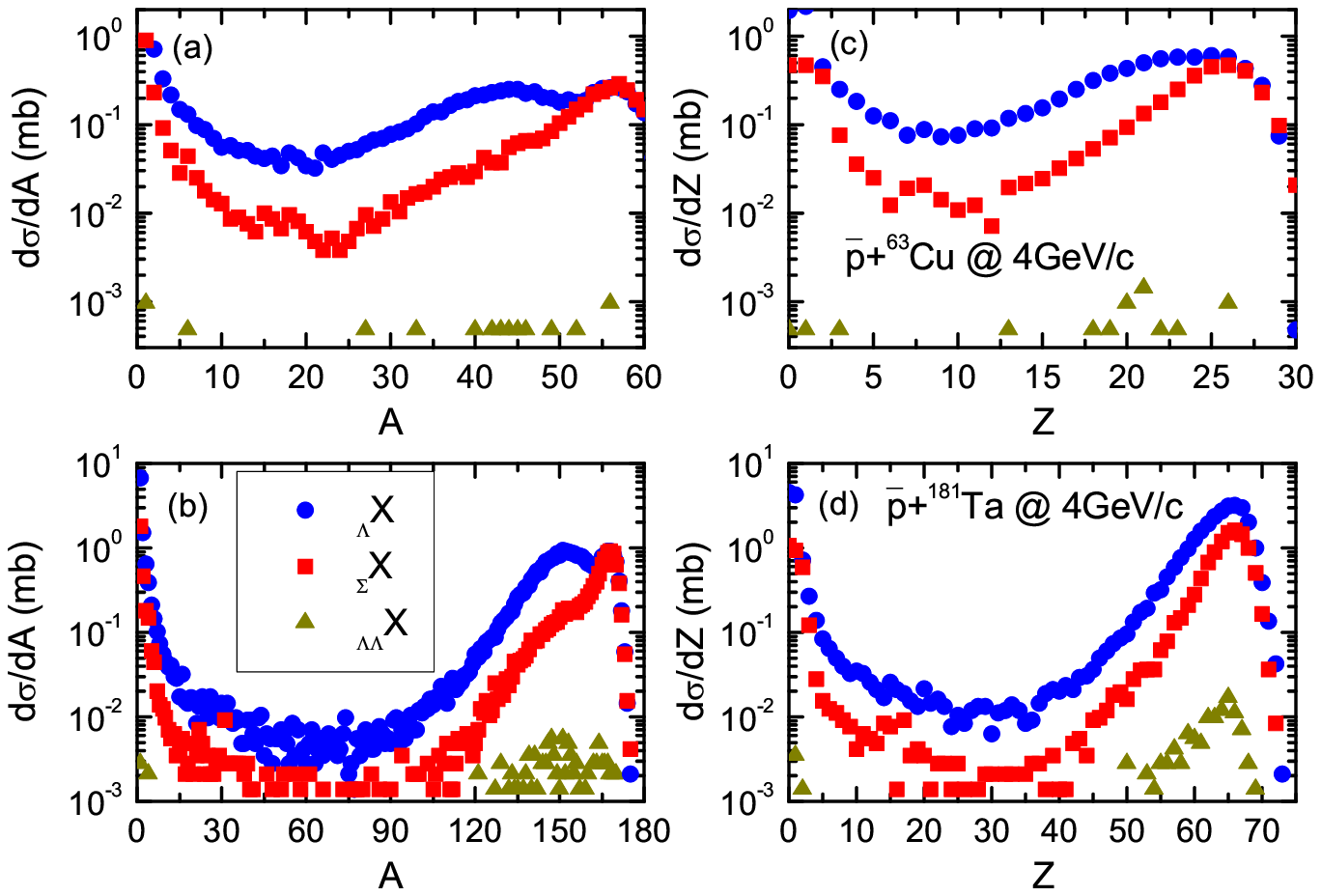}
\caption{(Color online) Hyperfragments with strangeness s=-1 ($_{\Lambda}$X and $_{\Sigma^{\pm}}$X) and s=-2 ($_{\Lambda\Lambda}$X) spectra as functions of atomic number (left panel) and charged number (right panel) in collisions of $\overline{p}$ on $^{63}$Cu and $^{181}$Ta at incident momentum of 4 GeV/c.}
\end{figure*}

The dynamics of the antiproton-nucleus collisions is more complicated in comparison to hadron (proton, $\pi$, $K$ etc) induced reactions and to heavy-ion collisions, in which the particles produced in the annihilation of the antiproton in a nucleus are coupled to the subsequent collisions with surrounding nucleons. The dynamics of antiproton-nucleus collisions is complicated, which is associated with the mean-field potentials of hadrons in nuclear medium, and also coupled to a number of reaction channels, i.e., the annihilation channels, charge-exchange reaction, elastic and inelastic collisions. A more localized energy deposition enables the secondary collisions available for producing hyperons. Hyperons produced in antiproton induced reactions can be captured in the potential of nucleon fragments to form hypernuclei. The dynamics of antiproton-nucleus collisions is complicated, which is associated with the mean-field potentials of hadrons in nuclear medium, and also with a number of reaction channels, i.e., the annihilation channels, charge-exchange reaction, elastic and inelastic collisions. The larger yields of strange particles in antiproton induced reactions are favorable to form hypernuclei in comparison to proton-nucleus and heavy-ion collisions. To understand the nuclear dynamics induced by antiprotons, several approaches have been proposed, such as the intranuclear cascade (INC) model \cite{Cu89}, kinetic approach \cite{Ko87}, Giessen Boltzmann-Uehling-Uhlenbeck (GiBUU) transport model \cite{La12,La15}, Statistical Multifragmentation Model (SMM) \cite{Bo95} and the Lanzhou quantum molecular dynamics (LQMD) approach \cite{Fe16c,Fe15b}. Particles production in antiproton induced reactions is significant in understanding the contributions of different reaction channels and the energy dissipation into the nucleus. Shown in Fig. 26 the temporal evolutions of pions, kaons, antikaons, hyperons and antihyperons in the reaction of $\overline{p}$+$^{40}$Ca at an incident momentum of 4 GeV/c. It is shown that the kaons are emitted immediately after the annihilation in collisions of antibaryons and baryons. The secondary collisions of pions and antikaons on nucleons retard the emissions and even a bump appears for antikaons in the reaction dynamics, i.e., $\pi N\rightarrow \Delta$, $\overline{K}N\rightarrow \pi Y$ etc, which lead to the production of hyperons. At the considered momentum above its threshold energy, e.g., the reaction $\overline{N}N\rightarrow \overline{\Lambda}\Lambda$ (p$_{\texttt{threshold}}$=1.439 GeV/c), production of hyperons are attributed from the direct reaction (annihilation and creation of quark pairs, $u\overline{u} (d\overline{d})\rightarrow s\overline{s}$) and also from the secondary collisions after annihilations in antibaryon-baryon collisions, i.e., meson induced reactions $\pi (\eta, \rho, \omega) N\rightarrow KY$ and strangeness exchange reaction $\overline{K}N\rightarrow \pi Y$. It is noticed that the antikaon yields are reduced for the heavy targets owing to more strong reabsorption reactions. However, the hyperon production is more pronounced with increasing the target mass, which is available for producing hypernucleus.

The fragmentation process of target nucleus induced by antiproton undergoes the explosive process (fast stage) in which the preequilibrium nucleons (light clusters) are emitted or the multifragments are produced after collisions between baryons and nucleons, and the decay process (slow stage) of the highly excited nucleus after the relative motion energy is deposited via the meson-nucleon collisions. The decay mechanism is determined by the excitation energy, i.e., the particle evaporation or fission dominating at low excitation energies (1-2 MeV/nucleon). The system is broken via multifragmentation when the local energy is close to the binding energy. More information on the hyperfragment formation in antiproton induced reactions is pronounced from the mass and charged number distributions. Direct production of hypernuclei with strangeness s=-2 (double $\Lambda$- hypernucleus) in heavy-ion collisions or by proton induced reactions are difficulty because of very mall cross sections, in particular for the heavy-mass region. Properties of the hypernuclei would be significant in understanding the $\Lambda$-$\Lambda$ and $\overline{\Lambda}$-nucleon interactions, which are not well understood up to now. The ($K^{-}$, $K^{+}$) reactions are used for producing the double hypernucleus $^{6}_{\Lambda\Lambda}$He \cite{Ta01}. The antiproton-nucleus collisions would be a chance for producing the s=-2 and s=1 hypernuclei. Shown in Fig. 27 is the mass and charged number distributions of hyperfragments with strangeness s=-1 ($_{\Lambda}$X and $_{\Sigma^{\pm}}$X) and s=-2 ($_{\Lambda\Lambda}$X) in collisions of $\overline{p}$ + $^{63}$Cu and $\overline{p}$ + $^{181}$Ta at the same of incident momentum of 4 GeV/c. The $\Lambda$-hyperfragments spread the whole isotope range with the larger yields. The maximal cross sections for s=-1 and s=-2 hyperfragments are at the levels of 1 $mb$ and 0.01 $mb$, respectively. The lower production yields of $\overline{\Lambda}$-hyperfragments at the level of 1 $\mu b$ are found \cite{Fe16c}. The hypernuclear physics with the antiproton beams are planing at the FAIR-PANDA (Antiproton Annihilation at Darmstadt) in the near future.

\section{Summary}

In this article, I reviewed the recent progress on the nuclear dynamics in heavy-ion collisions and in hadron induced reactions, in particular, on the topics of symmetry energy, in-medium properties of hadrons, fragmentation reactions and hypernucleus formation being stressed. The possible challenges and treatments in simulating nucleus-nucleus collisions with transport models are discussed. Further experiments are motivated at the facilities in the world such as CSR, HIAF, FRIB, Spiral2, FAIR etc. Within the framework of the LQMD transport model, the high-density symmetry energy, decay mode of highly excited nucleus, hyperon-nucleon and hyperon-hyperon interactions in dense nuclear matter, hypernuclide production with double strangeness are thoroughly investigated.

The isospin effect in Fermi-energy heavy-ion collisions is pronounced, in particular from the neck fragmentation. The ratios of neutron to proton yields from free nucleons and from light IMFs are sensitive to the stiffness of symmetry energy. The soft symmetry energy at the domain of subsaturation densities is constrained from the CHIMERA and MSU data. The mass splitting of neutron and proton in nuclear medium is obvious at the high beam energy. The optical potentials of pions, etas and kaons influence the structures of invariant spectra. The hadron production is coupled to the reaction channels and the transportation is related the mean-field potentials. The phase-space distributions of particles produced in hadron-hadron collisions are influenced by the potentials. A weakly repulsive $KN$ potential roughly being 25 MeV and the attractive $\overline{K}N$ potential being -100 MeV at  the normal density are obtained in comparison to the KaoS data. The pion-nucleon potential is not well constrained up to now, which impacts the pion energy spectra and associated with the properties of $\Delta$(1232) in nuclear medium. Parts of pions and etas are created at the subsaturation densities because of the reabsorption process. The rapidity and azimuthal angle cuts are necessary to extract the high-density symmetry energy from the yield ratios of $\pi^{-}/\pi^{+}$. Further experiments on the pion measurements are expected, such as HIRFL-CSR at Lanzhou, RIKEN-SAMURAI etc.

The nucleonic fragments and hyperfragments are investigated within the LQMD transport model associated with the GEMINI statistical code. Hyperfragments are formed in the region of  the narrower rapidity and lower kinetic energy in comparison with nucleonic fragments and hyperons. The hyperfragment yields formed in heavy-ion and proton (pion)-nucleus collisions are 10$-4$ times smaller in comparison to fragments. The nuclear reactions induced by K$^{-}$ and antiprotons are available for producing the $\Lambda$-hypernuclide. Moreover, the heavy-ion collisions, in particular with the neutron-rich beams, have advantage for creating the neutron-rich hypernuclide. Experiments on the strangeness physics and hypernucleus properties are planned at the HIAF in the future.

\end{document}